\documentstyle [preprint,pra,aps,12pt,tighten,floats,epsfig] {revtex}
\newcounter{bean}

\begin{document}

\begin{center}
~~~~~\\
{\large
{\bf DISSOCIATION, FRAGMENTATION AND FISSION
OF SIMPLE METAL CLUSTERS\footnote{
Contributed Chapter to the book {\bf Metal Clusters}, 
Edited by W. Ekardt (Wiley, New York, 1999) pp. 145-180.}}}\\
~~~~~\\
~~~~~\\
CONSTANTINE YANNOULEAS, UZI LANDMAN AND ROBERT N. BARNETT\\
{\it School of Physics, Georgia Institute of Technology}\\
{\it Atlanta, Georgia 30332-0430}
~~~~~\\
~~~~~\\
~~~~~\\
\end{center}
\section{Introduction}
Dissociation, fragmentation, and fissioning processes underly physical and 
chemical phenomena
in a variety of finite-size systems, characterized by a wide 
spectrum of energy scales,
nature of interactions, and characteristic spatial and temporal scales.
These include nuclear fission \cite{bm,pb}, unimolecular decay and
reactions in atoms and molecules \cite{two}, and more recently 
dissociation and fragmentation processes in atomic and molecular clusters 
\cite{brec,saun2,mart}.
Investigations of the energetics, mechanisms,
pathways, and dynamics of fragmentation processes provide ways and means for 
explorations of the structure, stability, excitations, and dynamics in the
many-body finite systems mentioned above, 
as well as they allow for comprehensive tests of
theoretical methodologies and conceptual developments, and have formed
active areas of fruitful research endeavors in nuclear physics, and
more recently in cluster science. 

Under the general title of dissociation and fragmentation \cite{note12}
processes in metal clusters, one usually distinguishes two 
classes of phenomena, i.e., (1) dissociation of neutral monomers and/or
dimers, and (2) fission. The physical processes in the first class are most 
often referrred to as evaporation of monomers and/or dimers, since they are
endothermic processes and are 
usually induced through laser heating of the cluster. The unimolecular 
equations associated with these processes are
\begin{equation}
M^+_N \longrightarrow M^+_{N-1} + M~,
\label{dism}
\end{equation}
for monomer separation, and
\begin{equation}
M^+_N \longrightarrow M^+_{N-2} + M_2~,
\label{disd}
\end{equation}
for dimer separation ($N$ denotes the number of atoms in the clusters
\cite{note11}). The parent clusters $M^+_N$ have been taken here as being
singly ionized, in order to conform with available experimental measurements
\cite{brec}. Fission on the other hand, is most often an exothermic process 
and is due to the Coulombic forces associated with excess charges on the 
cluster. It has been found that
the minimum excess charge required to induce fission is 2 elementary units 
(either positive or negative). In this case the doubly-charged parent 
cluster splits into two singly charged fragments, and the corresponding 
unimolecular equation can be written as
\begin{equation}
M^{2\pm}_N \longrightarrow M^{1\pm}_P + M^{1\pm}_{N-P},\;\;\; P=1,...,[N/2]~.
\label{disf}
\end{equation}
It needs to be emphasized that fragmentation through fission involves most 
often the overcoming of a fission barrier, while
momomer and dimer separation are barrierless processes \cite{brec}.

\subsection{Metal cluster fission and nuclear fission: Similarities
and differences} 

Multiply charged metallic clusters (M$_N^{Z+}$) are observable in mass
spectra if they exceed a critical size of stability $N_c^{Z+}$
(e.g. for $Z=2$, $N_c^{2+}=27$ for Na and $N_c^{2+}=20$ for K 
\cite{brec,brec4}).
For clusters with $N > N_c^{Z+}$, evaporation of neutral species is
the preferred dissociation channel, while, below the critical size,
fission into two charged fragments dominates (for $Z=2$, two singly charged
fragments emerge). Nevertheless, at low
enough temperature, such M$_N^{Z+}$ ($N < N_c^{Z+}$) clusters can
be metastable above a certain size $N_b^{Z+}$, because of the existence
of a fission barrier $E_b$ (for Na$_N^{2+}$ and K$_N^{2+}$, $N_b^{2+}=7$
\cite{barn,barn2}).

These observations indicate that fission of metal clusters occurs when
the repulsive Coulomb forces due to the accumulation of the excess
charges overcome the electronic binding (cohesion) of the cluster.
This reminds us immediately of the well-studied nuclear fission phenomenon
and the celebrated Liquid Drop Model (LDM)
according to which the binding
nuclear forces are expressed as a sum of volume and surface terms, and the
balance between the Coulomb repulsion and the increase in surface area
upon volume conserving deformations allows for an estimate of the
stability and fissility of the nucleus \cite{bw,ns}.

We note that for doubly charged metal clusters with $N \leq 12$ microscopic 
descriptions of energetics and dynamics of fission, based on first-principles
electronic-structure calculations in conjunction with molecular dynamics (MD) 
simulations, have been performed \cite{barn,barn2} (see section III.C.1 for
details). 
Several of the trends exhibited by the microscopic calculations (such as
influence of magic numbers, associated with electronic shell closing, on
fission energetics and barrier heights; predominance of an asymmetric
fission channel; double-humped fission-barrier shapes; shapes of deforming
clusters along the fission trajectory portraying two fragments connected 
through a stretching neck) suggest that appropriate adaptation 
of methodologies 
developed originally in the context of nuclear fission may provide a useful
conceptual and calculational framework for studies of systematics and patterns
of fission processes in metallic clusters.

In this context, it is useful 
to comment on the earliest treatments of pertinent nuclear processes,
i.e., fission \cite{bw,bm} and alpha radioactivity \cite{gamo,cg,pb}. 
Adaptation of the simple
one-center LDM to charged metallic clusters \cite{saun2}, involving calculation
of the Coulomb repulsive energy due to an excess charge localized at the
surface, yields a reduced LDM fissility parameter $\xi=(Z^2/N)/(Z^2/N)_{cr}$,
where $(Z^2/N)_{cr}=16 \pi r_s^3 \sigma/e^2$ with the surface energy per unit
area denoted by $\sigma$ and $r_s$ being the Wigner-Seitz radius 
(using bulk $r_s$ and $\sigma$ values,
$(Z^2/N)_{cr}=0.44$ and 0.39 for K$_N^{Z+}$ and Na$_N^{Z+}$, respectively).
Accordingly, a cluster is unstable for $\xi > 1$ (implying that for
K$_N^{2+}$ with $N \leq 9$ and Na$_N^{2+}$ with $N \leq 10$ barrierless
fission should occur) with the most favorable channel being the symmetric
one (i.e., when the two fragments have equal masses, which is only
approximately true for nuclear fission, and certainly not the case for small
metal clusters). For $0.351 < \xi < 1 $, the system is metastable (i.e.,
may fission in a process involving a barrier), and for $0 < \xi < 0.351$
the system is stable.

At the other limit, $\alpha$-radioactivity, which may be viewed as an
extreme case of (superasymmetric) fission, is commonly described as a process
where the fragments are formed (or as often said, preformed) before the
system reaches the top of the barrier (saddle point), and as a result the
barrier is mainly Coulombic \cite{pb}. We note here that asymmetric 
emission of heavier
nuclei is also known (e.g., $^{223}$Ra$\rightarrow^{14}$C$+^{209}$Pb, 
referred to as exotic or cluster radioactivity \cite{grei3,pric,grei2}),
and the barriers in these cases resemble the one-humped barrier of alpha
radioactivity and do not exhibit modulations due to shell effects \cite{grei2}.
We also remark that such $\alpha$-radioactivity-type (essentially Coulombic)
barriers have been proposed recently \cite{lope} for describing the
overall shape of the fission barriers in the case of metal clusters.

Although, several aspects of the simple LDM (e.g., competition between
Coulomb and surface terms) and the $\alpha$-particle, Coulombic 
model (e.g., asymmetric channels and a scission configuration close to the
location of the saddle of the multi-dimensional potential-energy surface)
are present in the fission of metal clusters, neither model is adequate
in light of the characteristic behavior revealed from the microscopic 
calculations and experiments. Rather, we find that proper treatments
of fission in these systems require consideration of shell effects
(for a recent experimental study that demonstrates the importance of
shell effects in metal-cluster fission, see Ref.\ [9b]).
While such effects are known to have important consequences in nuclear
fission (transforming the one-humped LDM barrier for symmetric fission into
a two-humped barrier \cite{nils2,pb}), their role in the case of metal 
clusters goes even further.
Indeed, as illustrated below (see section III.C.2) for the case of the magic 
Na$_{10}^{2+}$ (8 delocalized electrons), shell effects can be the largest 
contribution to the fission barrier, in particular in
instances when the LDM component exhibits no barrier (in this case the LDM
fissility $\xi > 1$). In this respect, Na$_{10}^{2+}$ is analogous to the case
of superheavy nuclei, which are believed \cite{ms} 
to be stabilized by the shell 
structure of a major shell closure at $Z_p=114$, $N_n=184$ 
($Z_p$ is the number of 
protons and $N_n$ is the number of neutrons; unfortunately such nuclei
have not been yet observed or synthesized artificially).

\subsection{Other decay modes in atomic and molecular clusters}

In this chapter, we will concentrate on the unimolecular processes 
in metal clusters described by Eqs. (\ref{dism}$-$\ref{disf}). However,
there is a variety of additional dissociation and fragmentation modes
in atomic and molecular clusters (see reviews in Ref.\ \cite{me}), 
which have been discovered experimentally or
anticipated theoretically; among them we mention:
\begin{enumerate}
\item
Unimolecular fission of triply and higher charged cationic simple metal
clusters \cite{mart,brec11,mart11};
\item
Metastability against electron autodetachment of multiply charged
{\it anionic\/} atomic clusters \cite{yl1,yl4,comp} and fullerenes 
\cite{yl4,comp,yl8};
\item
Fragmentation of cationic fullerenes via sequential evaporation
of carbon dimers \cite{mark11};
\item
Ultrarapid fragmentation of rare-gas clusters following excitation 
(involving excimer formation \cite{scha}) or ionization \cite{me}; 
\item
Multifragmentation phase transitions according to microcanonical 
thermodynamics of highly excited atomic clusters \cite{gros}; and
\item
Pathways and dynamics of dissociation and fragmentation of ionized 
Van-der-Waals and hydrogen-bonded molecular clusters 
\cite{me,blwa}.
\end{enumerate}

\subsection{Organization of the chapter}

In the following, we will present jellium-related theoretical approaches 
[specifically the Shell Correction Method (SCM) and variants thereof] 
appropriate for describing shell effects, energetics and decay pathways of 
metal-cluster fragmentation processes (both the monomer/dimer dissociation
and fission), which were inspired by the many similarities
with the physics of shell effects in atomic nuclei (section II). 
In section III, we will 
compare the experimental trends with the resulting theoretical SCM 
interpretations, and in addition we will discuss theoretical results from
first-principles MD simulations (section III.C.1). Section IV will discuss 
some latest insights concerning the importance of electronic-entropy and 
finite-temperature effects. Finally, section V will provide a summary.

\section{Theory of shape deformations}

In early applications of the jellium model, the shape of metal clusters 
was assumed in all instances 
to be spherical \cite{ekar,beck}, but soon it became apparent 
that the spherical symmetry was too restrictive \cite{clem,saun}.
Indeed clusters with open electronic shells (between the magic numbers
$N_e=2$, 8, 20, 40, 58, 92, etc...) are subjected to Jahn-Teller distortions
\cite{tell}.
By now it has been well established that a quantitative description of the 
underlying shell effects and of fragmentation phenomena (as well as of other 
less complicated phenomena such as Ionization and Vertical Electron Detachment)
requires a proper description of the deformed shapes of both parent and 
daughter clusters (of both precursor and final ionic or neutral product in 
the case of ionization and vertical electron detachment).

A most successful method for describing both deformation and shell
effects in simple metal
clusters (i.e., those that can be described by the jellium background model)
is the SCM, originally developed in the field of nuclear physics 
\cite{stru,pb}. In a series of recent publications 
\cite{yl1,yl4,yl8,yl2,yl3,yl5,yl6,yl7,yl9,yl10},
the SCM was further developed, adapted, and
applied in the realm of finite-size, condensed-matter nanostructures 
(i.e., metal clusters \cite{yl1,yl4,yl2,yl3,yl5,yl6,yl7},
but also multiply charged fullerenes \cite{yl8}, $^3$He clusters
\cite{yl9}, and metallic nanowires and nanoconstrictions 
\cite{yl10}).
Additionally, Refs.\ \cite{brac1,frau1,suga,vier} have used semiempirical 
versions (see below) of the SCM to study the shapes of neutral Na clusters 
\cite{brac1,frau1} and aspects of metal-cluster fission \cite{suga,vier}.

The SCM derives its justification from the local-density-approximation (LDA) 
functional theory and has been developed as a two-level method. 

At the microscopic level,
referred to as the LDA-SCM, the method has been shown to be a 
non-selfconsistent approximation to the Kohn-Sham (KS) $-$LDA approach
\cite{ks}. Apart from computational efficiency, 
an important physical insight provided by the LDA-SCM is
that the total KS-LDA energy $E_{\text{total}} (N)$ 
[or in another notation $E_{\text{KS}} (N)$] of a finite system of 
interacting delocalized electrons (or more generally of other fermions,
like nucleons or $^3$He atoms) can be divided into two contributions, i.e.,
\begin{equation}
E_{\text{total}} (N) =  \widetilde{E} (N) + \Delta E_{\text{sh}} (N)~,
\label{enshsm}
\end{equation}
where $\widetilde{E}$ is the part that varies smoothly as a function of the 
system size (e.g., the number, $N$ of atoms in a metal cluster), 
while $\Delta E_{\text{sh}} (N)$ is an oscillatory term accounting for the
shell effects; it 
arises from the discretization of the electronic states (quantum size effect).
$\Delta E_{\text{sh}} (N)$ is 
usually called a shell correction in the nuclear \cite{stru,bm}
and cluster \cite{yl1,yl4} literature.

Starting from the fundamental microscopic separation in Eq.\ (\ref{enshsm}), 
various semiempirical implementations (referred to as SE-SCM, see section 
II.B) of such a division consist of different approximate choices and methods 
for evaluating the two terms contributing to this separation. 

As an illustration of the physical content of Eq. (\ref{enshsm}) (which as 
well serves as a motivating example for the SCM),
we show in Fig.\ 1 the size-evolutionary pattern of the Ionization 
Potentials (IPs) of Na$_N$ clusters, which exhibits
\begin{figure}[t]
\centering\epsfig{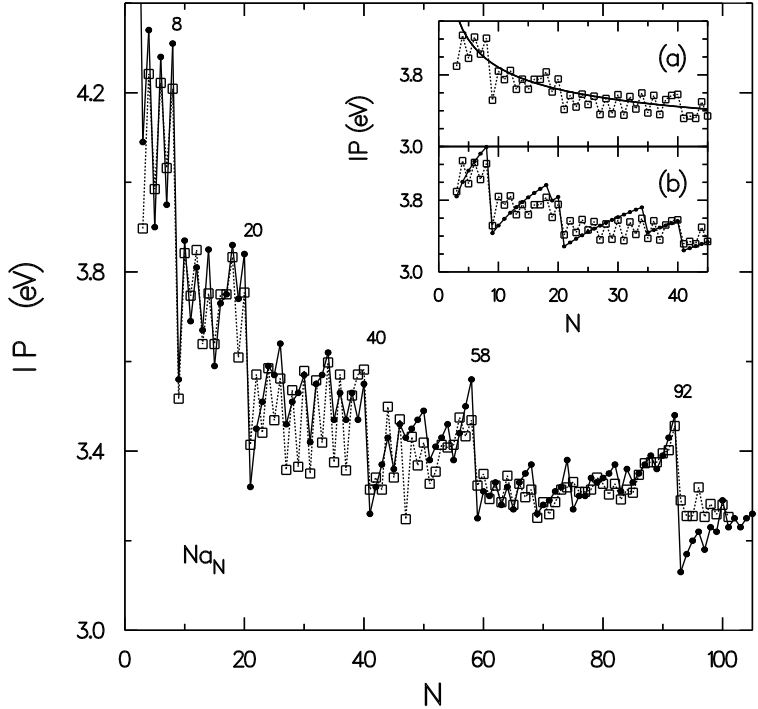}
\caption{
\small{
IPs of Na$_N$ clusters. Open squares: Experimental measurements 
\protect\cite{home,dehe2}. 
Solid circles: Theoretical IPs derived from the SCM assuming ellipsoidal 
(triaxial) deformations.
Inset (a): The solid line represents the smooth contribution to the theoretical
SCM IPs. 
Inset (b): The solid circles are the IPs derived from the SCM assuming 
spherical symmetry.}
}
\end{figure}
odd-even oscillations in the observed spectrum in addition to the major
features (major IP drops) at the magic numbers.
Theoretical calculations at three different levels are contrasted to the 
experimental observations, namely, a smooth
description of the pattern [Inset (a)], and two levels of shell-corrected
descriptions --- one assuming spherical symmetry [Inset (b)], and the other 
allowing for triaxial shape deformations [Fig.\ 1, main frame]. The progressive
improvement of the level of agreement between the experimental 
\cite{home,dehe2} and theoretical patterns is evident.

Below, we first outline the microscopic derivation of Eq.\ (\ref{enshsm}),
and subsequently we proceed with a presentation of the SE-SCM.

\subsection{Microscopic Foundation of Shell Correction Methods --
The LDA-SCM}

The LDA-SCM approach, which has been shown to yield results in excellent 
agreement with self-consistent KS-LDA calculations 
\cite{yl1,yl4}, is equivalent to a Harris functional \cite{harr}
approximation ($E_{\text{Harris}}[\rho^{\text{in}}]$, see below) 
to the KS-LDA total energy \cite{ks} 
$(E_{\text{KS}}[\rho_{\text{KS}}])$, 
with the input density
$\rho^{\text{in}}$ obtained through a variational
minimization of an extended Thomas-Fermi (ETF) energy functional,
$E_{\text{ETF}}[\rho]$.

The property of the non-selfconsistent Harris functional to yield total 
energies close to the KS-LDA ones is based on the following equality:
\begin{equation}
E_{\text{KS}}[\rho_{\text{KS}}] = E_{\text{Harris}}[\rho^{\text{in}}] +
O(\delta \rho^2)~,
\label{od2}
\end{equation}
where $\delta \rho = \rho_{\text{KS}} - \rho^{\text{in}}$.
Namely, the KS-LDA energy is, to second-order in $\delta \rho$, equal to
the Harris energy.  

Several recent publications have proven \cite{foul,finn,zare}
the validity of equation (\ref{od2}) in connection with the Harris
functional, which is often used in electronic structure calculations of 
molecules, surfaces, and other condensed-matter systems. 
We note that, in the context of nuclear physics, Strutinsky had earlier proven 
\cite{stru} the validity of Eq.\ (\ref{od2}), with the difference that he 
utilized the Hartree-Fock (HF) functional instead of the KS-LDA one. 
In the nuclear-physics literature, the HF version of Eq.\ (\ref{od2}) is 
referred to as the Strutinsky theorem.

Usually, in the Harris functional, 
the {\it input\/} density $\rho^{\text{in}}$ is 
taken as a superposition of site densities. Initially \cite{harr},
the site components of
the input density were not optimized. Later \cite{finn,zare}, it was 
realized that the results could be improved by variationally adjusting
the site components through a {\it maximization\/} of the Harris
functional itself. However, doing so adds the burden of a matrix 
diagonalization for obtaining the eigenvalues (see below) at each step of 
the variation. Our method differs from the Harris approach in that
the optimization of the input density is achieved by us through a variational
ETF method \cite{note55} 
(which does not require such a step-by-step matrix diagonalization).

The non-selfconsistent Harris functional is given by the following expression,
\begin{eqnarray}
E&&_{\text{Harris}} [\rho^{\text{in}}] = \nonumber \\
&&E_{\text{I}} + \sum_{i=1}^{\text{occ}} 
\epsilon_i^{\text{out}} 
 -  \int \! \left\{ \frac{1}{2} V_H [ \rho^{\text{in}} ({\bf r})] 
+ V_{\text{xc}} [ \rho^{\text{in}} ({\bf r})] \right\} 
\rho^{\text{in}} ({\bf r}) d{\bf r} 
+ \int \! {\cal E}_{\text{xc}} [ \rho^{\text{in}} ({\bf r})] d{\bf r}~,
\label{enhar}
\end{eqnarray}
where $V_H$ is the Hartree (electronic) repulsive potential, 
$E_{\text{I}}$ is the repulsive electrostatic energy of the ions, and
$E_{\text{xc}} [\rho] \equiv \int {\cal E}_{{xc}} [\rho] d {\bf r}$ 
is the exchange-correlation (xc) functional \cite{gunn}
[the corresponding xc potential is given as $V_{\text{xc}}({\bf r}) \equiv
\delta E_{\text{xc}} [\rho] / \delta \rho({\bf r})$].
$\epsilon_i^{\text{out}}$ are the eigenvalues
(non-selfconsistent) of the single-particle Hamiltonian,
\begin{equation}
\widehat{H} = - \frac{\hbar^2}{2m_e} \nabla^2 + V_{\text{in}}~,
\label{hin}
\end{equation}
with the mean-field potential given by
\begin{equation}
V_{\text{in}} 
[\rho^{\text{in}} ({\bf r})] =
V_H [\rho^{\text{in}} ({\bf r})]
+ V_{\text{xc}} [\rho^{\text{in}} ({\bf r})] 
+ V_I ({\bf r})~,
\label{mfpot}
\end{equation}
$V_I ({\bf r})$ being the attractive potential between the electrons and
ions.

The ETF-LDA energy functional, $E_{\text{ETF}} [\rho]$, 
is obtained by replacing the kinetic energy term,
$T[\rho]$, in the usual LDA functional, namely in the expression,
\begin{eqnarray}
E&&_{\text{LDA}}[\rho]= \nonumber \\
&& T[\rho] 
 + \int \left\{ \frac{1}{2} V_H [\rho({\bf r})]
 + V_I({\bf r}) \right\} \rho({\bf r})  d\/ {\bf r} 
 + \int {\cal E}_{\text{xc}} [\rho({\bf r})]d\/ {\bf r} + E_I~,
\label{enlda}
\end{eqnarray}
by the ETF kinetic energy, given to the 4th-order gradients as follows
\cite{hodg},
\begin{eqnarray}
&& T_{\text{ETF}}[\rho] =
\int t_{\text{ETF}}[\rho] d {\bf r}  \nonumber \\
&& = \frac{\hbar^2}{2m_e} 
\int \! \left\{
\frac{3}{5} (3\pi^2)^{2/3} \rho^{5/3} + 
\frac{1}{36} \frac{(\nabla \rho)^2}{\rho} +
\frac{1}{270} (3 \pi^2)^{-2/3} \rho^{1/3}  
\right. \nonumber \\
&& \times 
\left. \left[ 
\frac{1}{3} \left( \frac{\nabla \rho} {\rho}
\right)^4 - \frac{9}{8} \left( \frac{\nabla \rho} {\rho} \right)^2 
\frac{\Delta \rho} {\rho} +
 \left( \frac{\Delta \rho} {\rho} \right)^2 
\right] \right\}
d {\bf r}~.
\label{t4th}
\end{eqnarray}

We would like to remind the reader that the KS kinetic energy is of
course given by the expression
\begin{equation}
T_{\text{KS}}[\rho_{\text{KS}}] = 
\sum_{i=1}^{\text{occ}} < \phi_{{\text{KS}},i} | -\frac{\hbar^2}{2m_e} \nabla^2
|\phi_{{\text{KS}},i}>~,
\label{TKS}
\end{equation}
where the single-particle wave functions $\phi_{\text{KS},i} ({\bf r})$ are 
obtained from a self-consistent solution of the KS equations.

The optimal ETF-LDA total energy is obtained by minimization of
$E_{\text{ETF}} [\rho]$ 
with respect to the density. In our calculations, we use for the trial
densities parametrized profiles $\rho ({\bf r};\; \{\gamma_i\})$ 
\cite{brac2,yl1,yl4} with
$\{ \gamma_i \}$ as variational parameters (the ETF-LDA optimal density is
denoted as $\widetilde{\rho}$). The single-particle eigenvalues,
$\{\epsilon_i^{out}\}$, in Eq. (\ref{enhar})
are obtained then as the solutions to the single-particle Hamiltonian
of Eq.\ (\ref{hin}) with $V_{\text{in}}$ replaced by $V_{\text{ETF}}$
[given by Eq.\ (\ref{mfpot}) with $\rho^{\text{in}} ({\bf r})$
replaced by $\widetilde{\rho} ({\bf r})$].
Hereafter, these single-particle eigenvalues will be denoted by 
$\{ \widetilde{\epsilon}_i \}$.

In our approach,
the smooth contribution in the separation (\ref{enshsm}) of the total
energy is given by $E_{\text{ETF}} [\widetilde{\rho}]$, while
the shell correction, $\Delta E_{\text{sh}}$, is simply the difference
\cite{yl1,yl4}
\begin{eqnarray}
\Delta E_{\text{sh}} && = E_{\text{Harris}}[\widetilde{\rho}]
- E_{\text{ETF}} [\widetilde{\rho}] \nonumber \\
&& = \sum_{i=1}^{\text{occ}} \widetilde{\epsilon}_i -
\int \! \widetilde{\rho}({\bf r}) V_{\text{ETF}} ({\bf r}) d\/ {\bf r}
- T_{\text{ETF}} [\widetilde{\rho}]~.
\label{dsh}
\end{eqnarray}

\subsection{Semiempirical shell-correction method (SE-SCM)} 

\subsubsection{Methodology}

Rather than proceed with the microscopic route, Strutinsky proposed a method 
for the separation of the total energy into smooth and shell-correction terms 
[see Eq.\ (\ref{enshsm})] based on an averaging procedure.
Accordingly, a smooth part, $\widetilde{E}_{\text{sp}}$, is extracted out
of the sum of the single-particle energies 
$\sum_i^{\text{occ}} \widetilde{\epsilon}_i$
[or $\sum_i^{\text{occ}} \epsilon^{\text{out}}_i$, see Eq.\ (\ref{enhar})] 
by averaging them through an appropriate procedure. Usually, but not 
necessarily, one replaces the delta functions in the single-particle
density of states
by gaussians or other appropriate weighting functions. As a result, each
single-particle level is assigned an averaging occupation number
$\widetilde{f}_i$, and the smooth part $\widetilde{E}_{\text{sp}}$ is
formally written as
\begin{equation}
\widetilde{E}_{\text{sp}}=
\sum_i \widetilde{\epsilon}_i \widetilde{f}_i~.
\label{espoc}
\end{equation}

Consequently, the Strutinsky shell correction is given by
\begin{equation}
\Delta E_{\text{sh}}^{\text{Str}} = \sum_{i=1}^{\text{occ}} 
\widetilde{\epsilon}_i - \widetilde{E}_{\text{sp}}~.
\label{struav}
\end{equation}

The Strutinsky prescription
(\ref{struav}) has the practical advantage of using only the single-particle
energies $\widetilde{\epsilon}_i$, and not the smooth density 
$\widetilde{\rho}$. Taking advantage of this, the single-particle energies
can be taken as those of an external potential that empirically approximates
the self-consistent potential of a finite system. In the nuclear case,
a modified anisotropic three-dimensional harmonic oscillator has been used
successfully to describe the shell-corrections in deformed nuclei
\cite{bm,pb}.

The single-particle smooth part, $\widetilde{E}_{\text{sp}}$, however,
is only one component in the smooth contribution
$\widetilde{E}[\widetilde{\rho}]$, which needs to be added to the
shell correction term in order to yield the total energy, i.e.,
\begin{equation}
E_{\text{total}} \approx \Delta E_{\text{sh}}^{\text{Str}} + 
\widetilde{E}[{\widetilde{\rho}}]~.
\label{ehfsmsh}
\end{equation}

Strutinsky did not address the question of how to calculate microscopically
the smooth part $\widetilde{E}$ (which necessarily entails
specifying the smooth density ${\widetilde{\rho}}$). Instead he circumvented
this question by substituting for $\widetilde{E}$ the empirical
energies, $E_{\text{LDM}}$, of the nuclear liquid drop model, namely
he suggested that 
\begin{equation}
E_{\text{total}} \approx \Delta E_{\text{sh}}^{\text{Str}} +
E_{\text{LDM}}~.
\label{enhfld}
\end{equation}

In applications of Eq.\ (\ref{enhfld}),
the single-particle energies involved in the averaging [see Eqs.\ 
(\ref{espoc}) and (\ref{struav})] are commonly obtained as 
solutions of a Schr\"{o}dinger
equation with phenomenological one-body potentials. 
This last approximation has been very successful in describing
fission barriers and properties of strongly deformed nuclei using 
harmonic-oscillator-type or Wood-Saxon empirical potentials.

\subsubsection{Liquid-drop model for neutral and charged deformed clusters}

For neutral clusters, the LDM expresses \cite{brac2,suga,saun2} 
the {\it smooth\/}
part, $\widetilde{E}$, of the total energy as the sum of three contributions,
namely a volume, a surface, and a curvature term, i.e.,
\begin{eqnarray}
\widetilde{E} & = & E_{\text{vol}}+ E_{\text{surf}}+E_{\text{curv}} 
\nonumber \\
~ & = & A_v \int d \tau+ \sigma \int dS + A_c \int dS \kappa~,
\label{lqd}
\end{eqnarray}
where $d \tau$ is the volume element and $dS$ is the surface differential
element. The local curvature $\kappa$ is defined by the expression
$\kappa = 0.5 (R^{-1}_{\text{max}}+R^{-1}_{\text{min}})$, where
$R_{\text{max}}$ and $R_{\text{min}}$ are the two principal radii of
curvature at a local point on the surface of the 
jellium droplet (of a general shape)
which models the cluster. The corresponding coefficients
can be determined \cite{yl1,yl4,brac2} by fitting the 
ETF-LDA total energy for spherical shapes (see section II.A) to
the following parametrized expression as a function of the number, $N$,
of atoms in the cluster \cite{note1},
\begin{equation}
E_{\text{ETF}}^{\text{sph}} = \alpha_v N + \alpha_s N^{2/3} + \alpha_c
N^{1/3}~.
\label{npar}
\end{equation}
The following expressions relate \cite{note2} 
the coefficients $A_v$, $\sigma$, and
$A_c$ to the corresponding coefficients, ($\alpha$'s), in Eq.\ 
(\ref{npar}),
\begin{equation}
A_v = \frac{3}{4 \pi r_s^3} \alpha_v\; ; \;
\sigma = \frac{1}{4 \pi r_s^2} \alpha_s\; ; \;
A_c = \frac{1}{4 \pi r_s} \alpha_c~.
\label{conn}
\end{equation}

In the following, we will focus on the case of clusters with ellipsoidal
(triaxial) shapes.
In the case of ellipsoidal shapes the areal integral and the integrated
curvature can be expressed in closed analytical form with the help of
the incomplete elliptic integrals ${\cal F} (\psi,k)$ and ${\cal E} (\psi,k)$
of the first and second kind \cite{grad}, respectively. 
Before writing the formulas, we need to introduce some notations. 
Volume conservation must be employed, namely 
\begin{equation}
a^\prime b^\prime c^\prime /R_0^3 = abc =1~,
\label{volcons}
\end{equation}
where $R_0$ is the radius of a sphere with the same volume
($R_0=r_sN^{1/3}$ is taken to be the radius of the positive jellium
assuming spherical symmetry, $r_s$ being the corresponding Wigner-Seitz 
radius), and $a=a^\prime/R_0$, etc..., are the dimensionless semi-axes.
The eccentricities are defined through the dimensionless semi-axes as follows
\begin{eqnarray}
e_1^2 & = & 1 - (c/a)^2 \nonumber \\
e_2^2 & = & 1 - (b/a)^2 \nonumber \\
e_3^2 & = & 1 - (c/b)^2~. 
\label{ecc}
\end{eqnarray}
The semi-axes are chosen so that
\begin{equation}
a \geq b \geq c~.
\label{inqax}
\end{equation}

With the notation $\sin \psi = e_1$, $k_2= e_2/e_1$, and $k_3 = e_3/e_1$,
the relative (with respect to the spherical shape) surface and curvature 
energies are given \cite{hass} by 
\begin{equation}
\frac {E^{\text{ell}}_{\text{surf}}}{E^{\text{sph}}_{\text{surf}}} =
\frac{ab}{2} \left[ \frac{1-e_1^2}{e_1} {\cal F} (\psi, k_3)
+ e_1 {\cal E}(\psi, k_3) + c^3 \right]
\label{esurf}
\end{equation}
and
\begin{equation}
\frac { E^{\text{ell}}_{\text{curv}} } { E^{\text{sph}}_{\text{curv}} } =
\frac{bc}{2a} \left[ 1 + \frac{a^3}{e_1} 
\left( ( 1-e_1^2) {\cal F} (\psi, k_2) + e_1^2 {\cal E} (\psi, k_2) \right)
\right]~.
\label{ecurv}
\end{equation}

The change in the smooth part of the cluster total energy due to the excess
charge $\pm Z$ has been discussed for spherical clusters in Refs.\ 
\cite{yl1,yl4}. The result may be summarized as 
\begin{equation}
\Delta \widetilde{E}^{\text{sph}} (Z) =
\widetilde{E}^{\text{sph}}(Z)-\widetilde{E}^{\text{sph}}(0)=  \mp W Z +
\frac{Z(Z \pm 0.25)e^2}{2(R_0+\delta)},
\label{etznp}
\end{equation}
where the upper and lower signs correspond to negatively and positively
charged states, respectively, $W$ is the work function of the metal,
$R_0$ is the radius of the positive jellium assuming spherical symmetry,
and $\delta$ is a spillout-type parameter.

To generalize the above results to an ellipsoidal shape,
$\phi(R_0+\delta)$ $=$ $e^2/(R_0+\delta)$, which is the value of the
potential on the surface of a spherical conductor, needs to be replaced by
the corresponding expression for the potential on the surface of a 
conducting ellipsoid. The final result, normalized to 
the spherical shape, is given by the expression
\begin{equation}
\frac{\Delta \widetilde{E}^{\text{ell}}  (Z) \pm WZ}
{\Delta \widetilde{E}^{\text{sph}} (Z) \pm WZ} =
\frac{bc}{e_1} {\cal F}(\psi, k_2)~,
\label{ecoul}
\end{equation}
where the $\pm$ sign in front of $WZ$ corresponds to negatively and
positively charged clusters, respectively.

\subsubsection{The modified Nilsson potential for ellipsoidal shapes}

A natural choice for an external potential to be used for calculating
shell corrections with the Strutinsky method is an anisotropic,
three-dimensional oscillator with an additional ${\bf l}^2$ angular-momentum 
term for lifting the harmonic oscillator degeneracies \cite{nils}.
Such an oscillator model for approximating
the total energies of metal clusters, but without separating them into
a smooth and a shell-correction part in the spirit of Strutinsky' s approach,
had been used \cite{dehe2} with some success for calculating relative energy 
surfaces and deformation shapes of metal clusters. 
However, this simple harmonic oscillator model had serious limitations, 
since
i) the total energies were calculated by the expression
$\frac{3}{4} \sum_i^{\text{occ}} \widetilde{\epsilon}_i$, 
and thus did not compare with the total energies obtained from
the KS-LDA approach, and ii) the model could not be extended to the case of
charged (cationic or anionic) clusters. Thus absolute ionization potentials,
electron affinities, and fission energetics could not be calculated in this 
model. Alternatively, in our approach, we are making only a limited use of the 
external oscillator potential in calculating a modified Strutinsky shell 
correction. Total energies are evaluated by adding this shell correction to 
the smooth LDM energies (which incorporate xc contributions, since the LDM
coefficients are extracted via a comparison with total ETF-LDA energies,
or they are taken from experimental values).

In particular, a modified Nilsson Hamiltonian appropriate for metal
clusters \cite{clem,saun} is given by 
\begin{equation}
H_N = H_0 + U_0 \hbar \omega_0 ({\bf l}^2 -< {\bf l}^2>_n)~,
\label{hn}
\end{equation}
where $H_0$ is the hamiltonian for a three-dimensional anisotropic
oscillator, namely
\begin{eqnarray}
H_0 & = & -\frac{\hbar^2}{2m_e} \bigtriangleup +
       \frac{m_e}{2}(\omega_1^2 x^2 + \omega_2^2 y^2 +\omega_3^2 z^2)
 \nonumber \\
~& = & \sum_{k=1}^3 (a_k^\dagger a_k +\frac{1}{2}) \hbar \omega_k~.
\label{h0}
\end{eqnarray}

$U_0$ in Eq.\ (\ref{hn}) is a dimensionless parameter, which for occupied
states may depend on the effective principal quantum number $n=n_1+n_2+n_3$ 
associated with the major shells of any spherical-oscillator, $(n_1,n_2,n_3)$ 
being the quantum numbers specifying the single-particle levels 
of the hamiltonian $H_0$ (for clusters comprising up to 100 valence electrons, 
only a weak dependence on $n$ is found, see Table I in Ref.\ [40a]).
$U_0$ vanishes for values
of $n$ higher than the corresponding value of the last partially (or fully) 
filled major shell with reference to the spherical limit.

${\bf l}^2=\sum_{k=1}^3 l_k^2$ is a "stretched" angular momentum 
which scales to the ellipsoidal shape and is defined as follows,
\begin{equation}
l_3^2 \equiv (q_1p_2 - q_2 p_1)^2~,
\label{l3}
\end{equation}
(with similarly obtained expressions for $l_1$ and $l_2$ via a cyclic
permutation of indices) where the stretched position and 
momentum coordinates are defined via the corresponding natural coordinates,
$q^{\text{nat}}_k$ and $p^{\text{nat}}_k$, as follows,
\begin{equation}
q_k \equiv q^{\text{nat}}_k (m_e \omega_k/\hbar)^{1/2} = 
\frac{a_k^\dagger + a_k}{\sqrt{2}}~,~(k=1,2,3)~,
\label{q}
\end{equation}
\begin{equation}
p_k \equiv p^{\text{nat}}_k (1 /\hbar m_e \omega_k)^{1/2} = 
i\frac{a_k^\dagger - a_k}{\sqrt{2}}~,~(k=1,2,3)~.
\label{p}
\end{equation}

The stretched ${\bf l}^2$ is not a properly defined angular-momentum operator,
but has the advantageous property that it does not mix deformed states which
correspond to sherical major shells with different principal quantum numbers
$n=n_1+n_2+n_3$ (see, the Appendix in Ref.\ [40a]
for the expression of the matrix elements of ${\bf l}^2$).

The subtraction of 
the term $<{\bf l}^2>_n =n(n+3)/2$, where $<\; >_n$ denotes the expectation
value taken over the $nth$-major shell in spherical symmetry, 
guaranties that the average separation between major oscillator shells 
is not affected as a result of the lifting of the degeneracy.

The oscillator frequencies can be related to the principal semi-axes $a^\prime$, 
$b^\prime$, and $c^\prime$ [see, Eq.\ (\ref{volcons})] via the 
volume-conservation constraint and the requirement that the surface of the
cluster is an equipotential one, namely
\begin{equation}
\omega_1 a^\prime = \omega_2 b^\prime = \omega_3 c^\prime =
\omega_0 R_0~,
\label{omeg}
\end{equation}
where the frequency $\omega_0$ for the spherical shape (with radius $R_0$) 
was taken according to Ref.\ \cite{clem} to be
\begin{equation}
\hbar \omega_0 (N)= 
\frac{49 \; \mbox{eV bohr}^2}{r_s^2 N^{1/3}} 
\left[ 1 + \frac{t}{r_s N^{1/3}} \right]^{-2}~.
\label{omeg0}
\end{equation}
Since in this paper we consider solely monovalent elements, $N$ in Eq.\
(\ref{omeg0}) is the number of atoms for the family of clusters
M$_N^{Z \pm}$, $r_s$ is the Wigner-Seitz radius expressed in atomic units, 
and $t$ denotes the electronic spillout for the neutral cluster according
to Ref.\ \cite{clem}.

\subsubsection{Shell correction and averaging of single-particle spectra
for the modified Nilsson potential}

Usually $\widetilde{E}_{\text{sp}}$ 
[see Eqs.\ (\ref{espoc}) and (\ref{struav})] is calculated numerically 
\cite{nix}. 
However, a variation of the numerical Strutinsky averaging method consists in 
using the semiclassical partition function and in expanding it in powers of
$\hbar^2$. With this method, for the case of an anisotropic, fully triaxial 
oscillator, one finds \cite{bm,bhad} an analytical result, namely
\cite{note33}
\begin{equation}
\widetilde{E}_{\text{sp}}^{\text{osc}}  = 
\hbar (\omega_1 \omega_2 \omega_3)^{1/3}
\left( \frac{1}{4} (3N_e)^{4/3} +
\frac{1}{24} \frac{\omega_1^2 +\omega_2^2 +\omega_3^2}
{(\omega_1 \omega_2 \omega_3)^{2/3}} 
(3N_e)^{2/3} \right)~,
\label{harmav} 
\end{equation}
where $N_e$ denotes the number of delocalized valence electrons in the cluster.

In the present work, expression (\ref{harmav}) (as modified below) will be 
substituted for the average part
$\widetilde{E}_{\text{sp}}$ in Eq.\ (\ref{struav}), while the sum
$\sum_i^{\text{occ}} \widetilde{\epsilon}_i$ will be calculated numerically by
specifying the occupied single-particle states of the modified Nilsson
oscillator represented by the hamiltonian (\ref{hn}).

In the case of an isotropic oscillator, not only the smooth
contribution, 
$\widetilde{E}_{\text{sp}}^{\text{osc}}$, but also the Strutinsky shell 
correction (\ref{struav}) can be specified analytically \cite{bm}
with the result
\begin{equation}
\Delta E^{\text{Str}}_{\text{sh,0}}(x) =
\frac{1}{24} \hbar \omega_0 (3N_e)^{2/3} (-1 + 12 x (1-x))~,
\label{harmsh}
\end{equation}
where $x$ is the fractional filling of the highest partially filled 
harmonic-oscillator major shell. For a filled shell 
($x=0$ or 1), $\Delta E^{\text{Str}}_{\text{sh,0}}(0) = - \frac{1}{24}
\hbar \omega_0 (3N_e)^{2/3}$, instead of the essentially vanishing value 
as in the case of the ETF-LDA defined shell correction (cf. Fig.\ 1
of Ref.\ [40a]).
To adjust
for this discrepancy, we add $-\Delta E^{\text{Str}}_{\text{sh,0}}(0)$ to 
$\Delta E^{\text{Str}}_{\text{sh}}$ calculated through Eq.\ (\ref{struav})
for the case of open-shell, as well as closed-shell clusters.

\subsubsection{Overall procedure}

We are now in a position to summarize the calculational procedure for the
SE-SCM in the case of deformed clusters, which consists of the following steps:

\begin{enumerate}

\item Parametrize results of ETF-LDA calculations for spherical neutral
      jellia according to Eq.\ (\ref{npar}).

\item Use above parametrization (assuming that parameters per differential 
      element of volume, surface, and integrated curvature are shape 
      independent) in Eq.\ (\ref{lqd}) to calculate the liquid-drop energy 
      associated with neutral clusters, and then add to it the charging
      energy according to Eq.\ (\ref{ecoul}) to determine the total LDM
      energy $\widetilde{E}$ (available experimental values for $\sigma$
      and $W$ can also be used).
      
\item Use Equations (\ref{hn}) and (\ref{h0}) for a given deformation
      [i.e., $a^\prime$, $b^\prime$, $c^\prime$, or equivalently
      $\omega_1$, $\omega_2$, $\omega_3$, see Eq.\ (\ref{omeg})] to solve for 
      the single-particle spectrum ($\widetilde{\epsilon}_i$).
      
\item Evaluate the average, 
      $\widetilde{E}_{\text{sp}}$, of the single-particle spectrum
      according to Eq.\ (\ref{harmav}) and subsequent remarks.

\item Use the results of steps 3 and 4 above to calculate the shell correction
      $\Delta E_{\text{sh}}^{\text{Str}}$ according to Eq.\ (\ref{struav}).

\item Finally, calculate the total energy $E_{\text{total}}$ as the sum of the
      liquid-drop contribution (step 2) and the shell correction (step 5),
     namely $E_{\text{total}}=\widetilde{E}+\Delta E_{\text{sh}}^{\text{Str}}$.

\end{enumerate}
      
The optimal ellipsoidal geometries for a given cluster M$^{Z\pm}_N$, either 
neutral or charged, are determined by systematically varying the distortion 
(namely, the parameters $a$ and $b$) in order to locate the global minimum
of the total energy $E_{\text{total}}(N,Z)$ (for the global minima and 
equilibrium shapes of neutral Na$_N$ clusters according to the
ellipsoidal model in the range $3 \leq N \leq 60$, see
Fig.\ 22 of Ref.\ [40a]).

\subsubsection{Asymmetric two-center oscillator model for fission}

Naturally, the one-center Nilsson oscillator is not the most appropriate
empirical potential for describing binary fission, which involves the gradual 
emergence of two separate fragments. A better choice is the asymmetric 
two-center oscillator model (ATCOM). According to the ATCOM approach,
the single-particle levels, associated with both the initial
one-fragment parent and the separated daughters emerging from binary cluster
fission, are determined by the following single-particle hamiltonian
\cite{grei,must},
\begin{equation}
H=T + \frac{1}{2} m_e \omega^2_{\rho i} \rho^2
    + \frac{1}{2} m_e \omega^2_{z i} (z-z_i)^2 + V_{\text{neck}}(z)
    + U({\bf l}_i^2)~,
\label{hsp}
\end{equation}
where $i=1$ for $z<0$ (left) and $i=2$ for $z>0$ (right).

This hamiltonian is axially symmetric along the $z$ axis.
$\rho$ denotes the cylindrical coordinate perpendicular to the symmetry axis
\cite{note43}.
The shapes described by this Hamiltonian are those of two semispheroids
(either prolate or oblate)
connected by a smooth neck [which is specified by the term 
$V_{\text{neck}}(z)$].
$z_1 < 0$ and $z_2 > 0$ are the centers of these semispheroids. For the smooth
neck, the following 4th-order expression \cite{must} was adopted, namely
\begin{equation}
V_{\text{neck}}(z) = \frac{1}{2} m_e \xi_i \omega^2_{z i} (z-z_i)^4 
                     \theta(|z|-|z_i|)~,
\label{vneck}
\end{equation}
where $\theta(x)=0$ for $x>0$ and $\theta(x)=1$ for $x<0$ and 
$\xi_i=-1/2z_i^2$.

The frequency $\omega_{\rho i}$ in Eq. (\ref{hsp}) must be $z$-dependent in 
order to interpolate smoothly between the values $\omega^\circ_{\rho i}$ of the
lateral frequencies associated with the left ($i=1$) and right ($i=2$) 
semispheroids, which are not equal in asymmetric cases. The frequencies 
$\omega^\circ_{\rho i}$ $(i=1,2)$ characterize the latteral harmonic
potentials associated with the two semispheroids outside the neck region.
In the implementation of such an interpolation, we folllow 
Ref.\ \cite{must}.

The angular-momentum dependent term $U({\bf l}_i^2)$, where ${\bf l}_1$
and ${\bf l}_2$ are pseudoangular momenta with respect to the left and right
centers $z_1$ and $z_2$, is a direct generalization of the corresponding term
familiar from the one-center Nilsson potential (e.g., see Ref.\ [40a]).
Its function is to lift the usual harmonic-oscillator
degeneracies for different angular momenta, that is, for a spherical shape
the $1d-2s$ degeneracy is properly lifted into a $1d$ shell that is lower
than the $2s$ shell (for the parameters entering into this term, which ensure
a proper transition from the case of the fissioning cluster to that of
the separated two fragments, we have followed Ref.\ \cite{must}).

The cluster shapes associated with the spatial-coordinate-dependent 
single-particle potential
$V(\rho,z)$ in the hamiltonian (\ref{hsp}) (i.e., the second, thrid, and
fourth terms) are determined by the assumption 
that the cluster surface coincides with an equipotential
surface of value $V_0$, namely, from the relation $V(\rho,z)=V_0$.
Subsequently, one 
solves for $\rho$ and derives the cluster shape $\rho=\rho(z)$. For the
proper value of $V_0$, we take the one associated with a spherical shape
containing the same number of atoms, $N$, as the parent cluster, namely,
$V_0=\frac{1}{2} m_e \omega_0^2 R^2$, 
where $\hbar \omega_0 = 49 r_s^{-2} N^{-1/3}$ eV, $R=r_s N^{1/3}$,
and $r_s$ is the Wigner-Seitz radius in atomic units (monovalent metals
have been assumed). Volume conservation is implemented by
requiring that the volume enclosed by the fissioning cluster surface 
(even after separation) remains equal to $4\pi R^3/3$.

The cluster shape in this parametrization is specified by 
four independent parameters. We take them
to be:  the separation $d=z_2-z_1$ of the semispheroids; the asymmetry
ratio $q_{as}=\omega^\circ_{\rho2}/\omega^\circ_{\rho1}$; and the
deformation ratios for the left (1) and right (2) semispheroids 
$q_i=\omega_{zi}/\omega^\circ_{\rho i}$ $(i=1,2)$.

The single-particle levels of the hamiltonian in Eq.\ (\ref{hsp}) are obtained
by numerical diagonalization in a basis consisting of the eigenstates of
the following auxiliary hamiltonian:
\begin{equation}
H_0=T + \frac{1}{2} m_e \overline{\omega}_\rho^2 \rho^2
      + \frac{1}{2} m_e \omega_{zi}^2 (z-z_i)^2~,
\label{h00}
\end{equation}
where 
$\overline{\omega}_\rho$ is the arithmetic average of
$\omega^\circ_{\rho1}$ and $\omega^\circ_{\rho2}$.
The eigenvalue problem specified by the auxiliary hamiltonian
(\ref{h00}) is separable in the cylindrical variables $\rho$ and $z$.
The general solutions in $\rho$ are those of a two-dimensional
oscillator, while in $z$ they can be expressed through the parabolic
cylinder functions \cite{abra}.
The matching conditions at $z=0$ for the left and
right domains yield the $z$-eigenvalues and the associated eigenfunctions
\cite{grei}.

Having obtained the single-particle spectra, the empirical
shell correction (in the spirit of Strutinsky's method \cite{stru}), 
$\Delta E_{\text{sh}}^{\text{Str}}$, is determined from Eq.\ (\ref{struav}).

The single-particle average, $E^{\text{Str}}_{\text{av}}$
[i.e., $\widetilde{E}_{\text{sp}}$ in Eq.\ (\ref{struav})], is calculated 
\cite{jenn} through an $\hbar$ expansion of the 
semiclassical partition function introduced by Wigner and Kirkwood
(see references in Ref.\ \cite{jenn}). For general-shape potentials, this
last method amounts \cite{jenn} to eliminating the semiclassical Fermi energy
$\widetilde{\lambda}$ from the set of the following two 
equations
\begin{equation}
N_e = \frac{1}{3\pi^2} \left( \frac{2m_e}{\hbar^2} \right)^{3/2}
    \int^{ {\bf r}_{\tilde{\lambda}} } d{\bf r}
    \left[ (\widetilde{\lambda}-V)^{3/2}
    -\frac{1}{16} \frac{\hbar^2}{2m_e} (\widetilde{\lambda}-V)^{-1/2} 
    \nabla^2 V \right]~,
\label{smne}
\end{equation}
and
\begin{eqnarray}
E^{\text{Str}}_{\text{av}} 
& = &\frac{1}{3\pi^2} \left( \frac{2m_e}{\hbar^2} \right)^{3/2}
    \int^{{\bf r}_{\tilde{\lambda}}} d{\bf r}
    \left(
    \left[ \frac{3}{5} (\widetilde{\lambda}-V)^{5/2}
           +V(\widetilde{\lambda}-V)^{3/2}  \right] \right. \nonumber \\
  & &  \left. + \frac{1}{16} \frac{\hbar^2}{2m_e} 
     \left[ ({\widetilde{\lambda}}-V)^{1/2} \nabla^2 V
       - V({\widetilde{\lambda}}-V)^{-1/2} \nabla^2 V \right]
     \right)~,
\label{smaven}
\end{eqnarray}
where $N_e$ is the total number of delocalized valence electrons, 
and $V(\rho,z)$ is the potential in the single-particle
hamiltonian of Eq.\ (\ref{hsp}). The domain of integration is demarcated by the
classical turning point ${\bf r}_{\tilde{\lambda}}$, such that 
$V({\bf r}_{\tilde{\lambda}})={\widetilde{\lambda}}$.

Finally, from the liquid-drop-model contributions, we retain
the two most important ones, namely the surface contribution and the 
Coulomb repulsion. To determine the surface contribution, we calculate
numerically the area of the 
surface of the fissioning cluster shape, $\rho=\rho(z)$, 
and multiply it by a surface-tension coefficient specified via an
ETF-LDA calculation for spherical jellia
\cite{yl1,yl2} (or even from experimental values).
The Coulomb repulsion is calculated numerically
using the assumption of a classical conductor, namely the excess 2 units
of positive charge are assumed to be distributed over the surface of the
fissioning cluster, and in addition each of the fragments carries one unit
of charge upon separation (for a more elaborate 
application of the LDM to triaxially deformed ground states of neutral and
charged metal clusters described via a one-center shape parametrization, 
see our discussion in section II.B.2 in connection with Eqs. 
(\ref{lqd}-\ref{ecoul}) and Ref.\ \cite{yl3}).

As a result, the total energy $E_{\text{total}}$ for a specific fission
configuration is given by
\begin{equation}
E_{\text{total}}=E_{\text{LDM}}+\Delta E_{\text{sh}}^{\text{Str}}=
E_S + E_C + \Delta E_{\text{sh}}^{\text{Str}}~,
\label{et}
\end{equation}
where $E_S$ and $E_C$ are the surface and Coulomb terms, respectively.

\section{Experimental trends and theoretical interpretation}

In the following, we describe applications of the SE-SCM approach to systematic
investigations of the effects of shape deformations on the 
energetics of fragmentation processes 
of metal clusters \cite{yl4,yl3,yl5}, and to studies of deformations
and barriers in fission of charged metal clusters \cite{yl7}. We mention that,
in addition, Strutinsky calculations using phenomenological potentials have 
been reported for the case of neutral sodium clusters assuming axial symmetry 
in Refs.\ \cite{brac1,frau1,bulg}, and for the case of fission in Refs.\ 
\cite{suga,vier}.

\subsection{Electronic shell effects in monomer and dimer separation energies}

Monomer and dimer separation energies associated with the unimolecular 
reactions K$_N^+ \rightarrow $ K$_{N-1}^+ + $K, 
K$_N^+ \rightarrow $ K$_{N-2}^+ + $K$_2$, 
and Na$_N^+ \rightarrow $ Na$_{N-1}^+ + $Na can be calculated as
follows
\begin{equation}
D^+_{1,N}=E_{\text{total}} ({\cal Z}=+1,N-1)+E_{\text{total}}({\cal Z}=0,N=1) 
-E_{\text{total}}({\cal Z}=+1,N)~,
\label{d1n}
\end{equation}
and
\begin{equation}
D^+_{2,N}=E_{\text{total}} ({\cal Z}=+1,N-2)+E_{\text{total}}({\cal Z}=0,N=2) 
-E_{\text{total}}({\cal Z}=+1,N)~,
\label{d2n}
\end{equation}
where ${\cal Z} = \pm Z$ ($Z$ being the excess positive
or negative charge in absolute units).

\begin{figure}[t]
\centering\epsfig{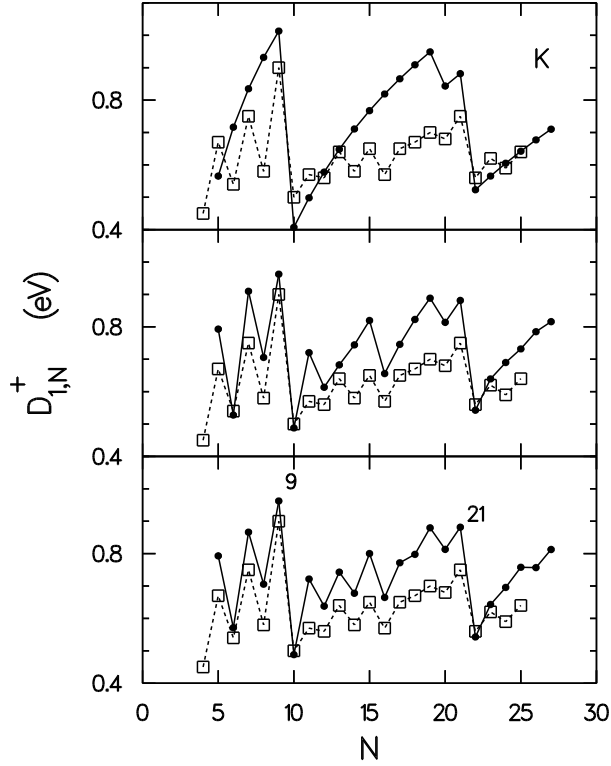}
\caption{
\small{
Monomer separation energies, $D_{1,N}^+$ [see Eq. (\protect\ref{d1n})],
from singly cationic K$^+_N$ clusters in the range 
$ 5 \leq N \leq  27 $.
Solid dots: Theoretical results derived from the SE-SCM method.
Open squares: Experimental measurements from Ref.\protect\ 
\protect\onlinecite{brec1}.
Top panel: The spherical model compared to experimental data.
Middle panel: The spheroidal (axially symmetric) model compared to 
experimental data.
Lower panel: The ellipsoidal (triaxial) model compared to experimental data.
}}
\end{figure}
The theoretical results for $D^+_{1,N}$ and $D^+_{2,N}$ for potassium are 
displayed in Fig.\ 2 and Fig.\ 3, respectively, and are compared to the
experimental measurements \cite{brec1}. The theoretical and experimental 
\cite{brec2} results for $D^+_{1,N}$ in the case of sodium are displayed 
in Fig.\ 4 (bottom panel). 
An inspection of all three figures leads to the same conclusion 
as in the case of IPs and electron affinities
(see Fig.\ 1 and Ref.\ [40a]), 
i.e., that results obtained via 
\begin{figure}[t]
\centering\epsfig{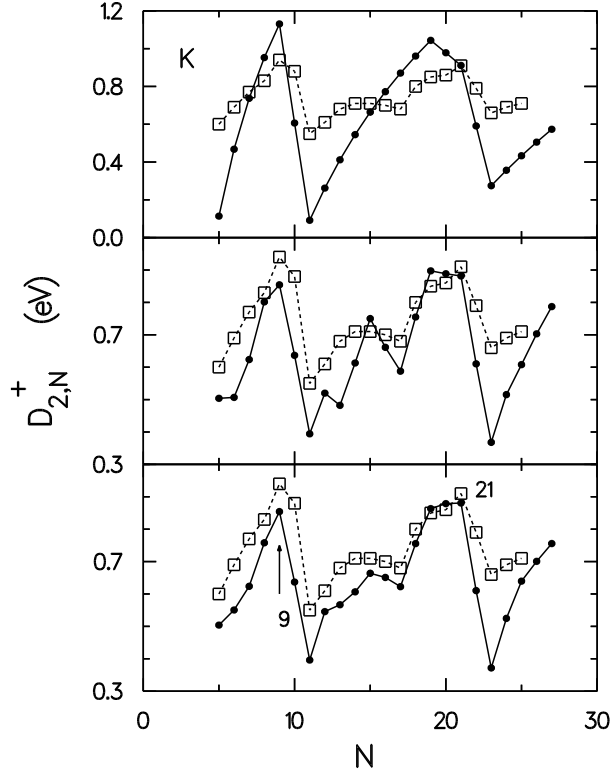}
\caption{
\small{
Dimer separation energies, $D_{2,N}^+$ [see Eq. (\protect\ref{d2n})], 
from singly cationic K$^+_N$ clusters in the range 
$ 5 \leq N \leq  27 $.
Solid dots: Theoretical results derived from the SE-SCM method.
Open squares: Experimental measurements from Ref.\protect\ 
\protect\onlinecite{brec1}.
Top panel: The spherical model compared to experimental data.
Middle panel: The spheroidal model compared to experimental data.
Lower panel: The ellipsoidal model compared to experimental data.}}
\end{figure}
\begin{figure}[t]
\centering\epsfig{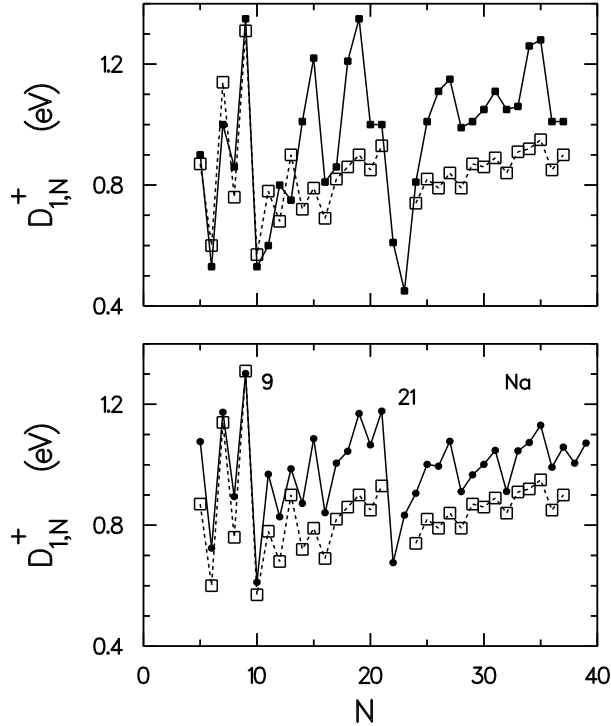}
\caption{
\small{
Monomer separation energies, $D_{1,N}^+$ [see Eq. (\protect\ref{d1n})], 
from singly cationic Na$^+_N$ clusters in the range 
$ 5 \leq N \leq  39 $.
Open squares: Experimental measurements from Ref.\protect\ 
\protect\cite{brec2}.
Solid dots (Bottom panel): Theoretical results derived from the SE-SCM method
in the case of triaxial deformations.
Solid squares (Top panel): Theoretical results according to the KS-LDA 
spheroidal calculations of Ref.\ \protect\cite{ekar8}.}}
\end{figure}
calculations restricted to spherical shapes 
compare rather poorly with the experiment, that improvement is evident when 
spheroidal (axially symmetric) deformations are considered, and that the 
agreement between theory and experiment becomes detailed when triaxiality 
(i.e., ellipsoidal shapes) is taken into consideration.
The feature of the appearance of strong odd-even alternations for $N=12-15$
together with a well-defined quartet in the range $N=16-19$ is present
in the experimental monomer separation energies of both potassium and sodium 
clusters, and theoretically it can be accounted for only after the inclusion 
of triaxial deformations. 

We note that in the case of dimer separation energies (Fig.\ 3) the odd-even
alternations cancel out. Parents with closed shells or subshells correspond
to maxima, while daughters with closed shells or subshells are associated
with minima (e.g., the triplets $N=9-11$, or $N=15-17$).

We also include for comparison results obtained by KS-LDA calculations 
\cite{ekar8} for deformed Na$_N$ clusters restricted to spheroidal (axial)
symmetry (Fig. 4, top panel). As expected, except for very small clusters
($N < 9$), these results do not exhibit odd-even oscillations. In addition,
significant discrepancies between the calculated and experimental results are
evident, particularly pertaining to the amplitude of oscillations at shell
and subshell closures.

\subsection{Electronic shell effects in fission energetics}

Fission of doubly charged metal clusters, M$^{2\pm}_N$, has attracted 
considerable attention in the last few years. 
LDA calculations for fission energetics have usually been restricted to 
spherical jellia for both parent and 
daughters, \cite{balb,lope} with the exception of molecular-dynamical 
calculations for sodium \cite{barn} and potassium \cite{barn2} clusters with 
$N \leq 12$. 
We present here systematic calculations for the dissociation 
energies $\Delta_{N,P}$ of the fission processes $\text{K}^{2+}_N \rightarrow 
\text{K}^+_{P} + \text{K}^+_{N-P}$, as a function of the fission channels $P$.

We have calculated the dissociation energies
\begin{equation}
\Delta_{N,P} = E_{\text{total}} ({\cal Z}=+1,P)+E_{\text{total}}
({\cal Z}=+1,N-P) - E_{\text{total}}({\cal Z}=+2,N)~,
\label{delt}
\end{equation}
\begin{figure}[t]
\centering\epsfig{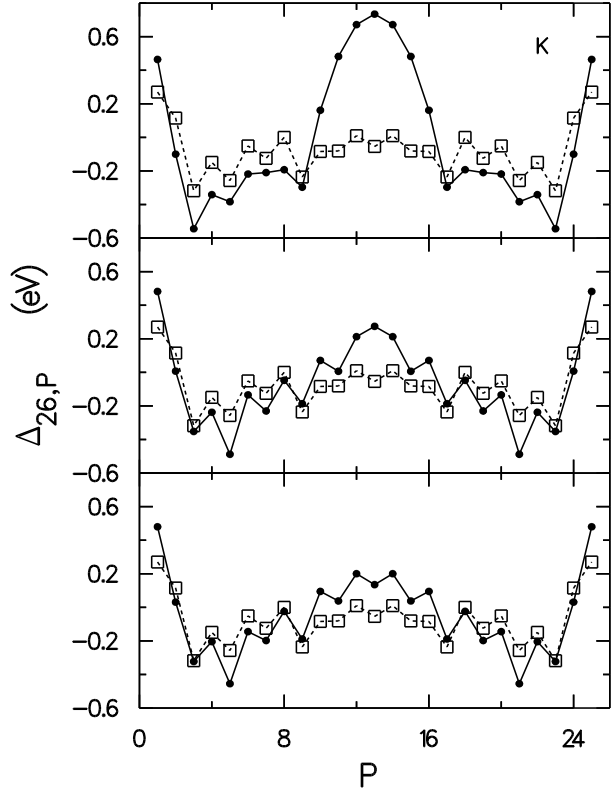}
\caption{
\small{
Fission dissociation energies, $\Delta_{26,P}$ [see Eq. (\protect\ref{delt})], 
for the doubly cationic K$^{2+}_{26}$ cluster 
as a function of the fission channels $P$.
Solid dots: Theoretical results derived from the SE-SCM method.
Open squares: Experimental measurements from Ref.\protect\ 
\protect\onlinecite{brec3}.
Top panel: The spherical model compared to experimental data.
Middle panel: The spheroidal model compared to experimental data.
Lower panel: The ellipsoidal model compared to experimental data.
}}
\end{figure}
\begin{figure}[t]
\centering\epsfig{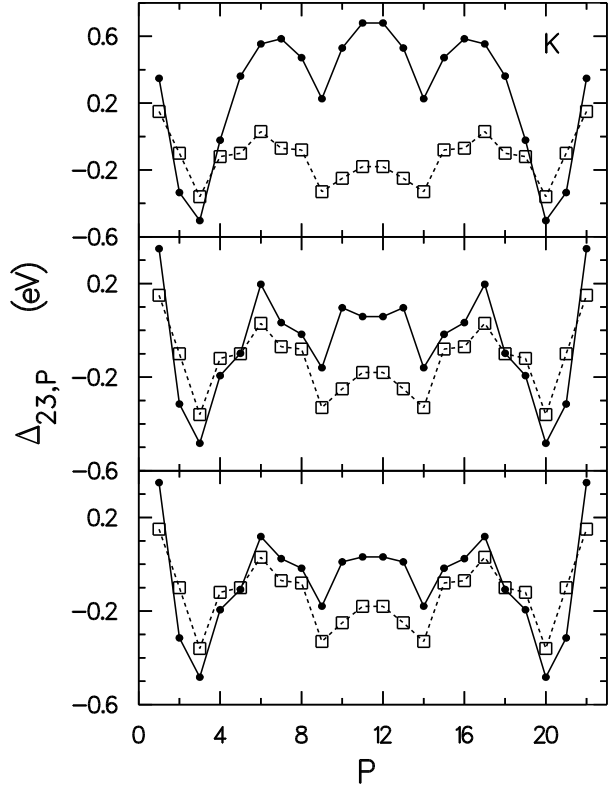}
\caption{
\small{
Fission dissociation energies, $\Delta_{23,P}$ [see Eq. (\protect\ref{delt})], 
for the doubly cationic K$^{2+}_{23}$ cluster 
as a function of the fission channels $P$.
Solid dots: Theoretical results derived from the SE-SCM method.
Open squares: Experimental measurements from Ref.\protect\ 
\protect\onlinecite{brec3}.
Top panel: The spherical model compared to experimental data.
Middle panel: The spheroidal model compared to experimental data.
Lower panel: The ellipsoidal model compared to experimental data.
}}
\end{figure}
\begin{figure}[t]
\centering\epsfig{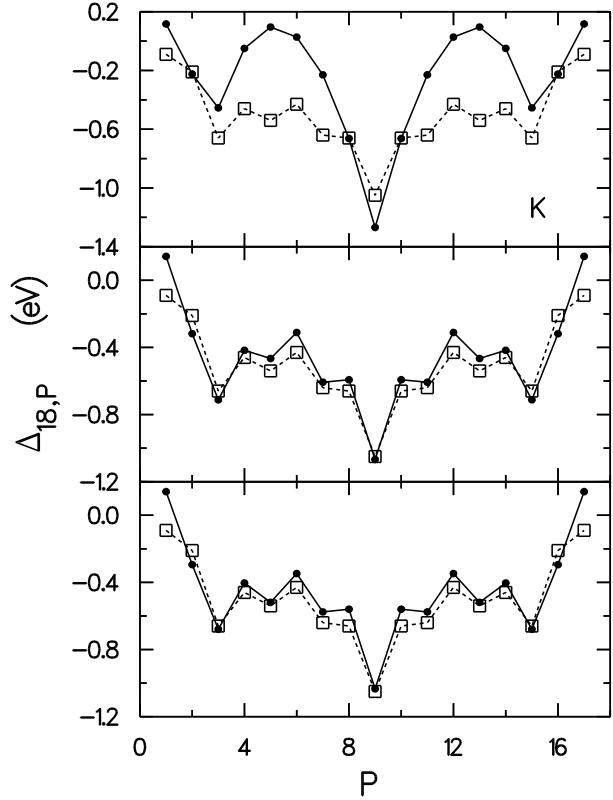}
\caption{
\small{
Fission dissociation energies, $\Delta_{18,P}$ [see Eq. (\protect\ref{delt})],
for the doubly cationic K$^{2+}_{18}$ cluster 
as a function of the fission channels $P$.
Solid dots: Theoretical results derived from the SE-SCM method.
Open squares: Experimental measurements from Ref.\protect\ 
\protect\onlinecite{brec3}.
Top panel: The spherical model compared to experimental data.
Middle panel: The spheroidal model compared to experimental data.
Lower panel: The ellipsoidal model compared to experimental data.
}}
\end{figure}
\begin{figure}[t]
\centering\epsfig{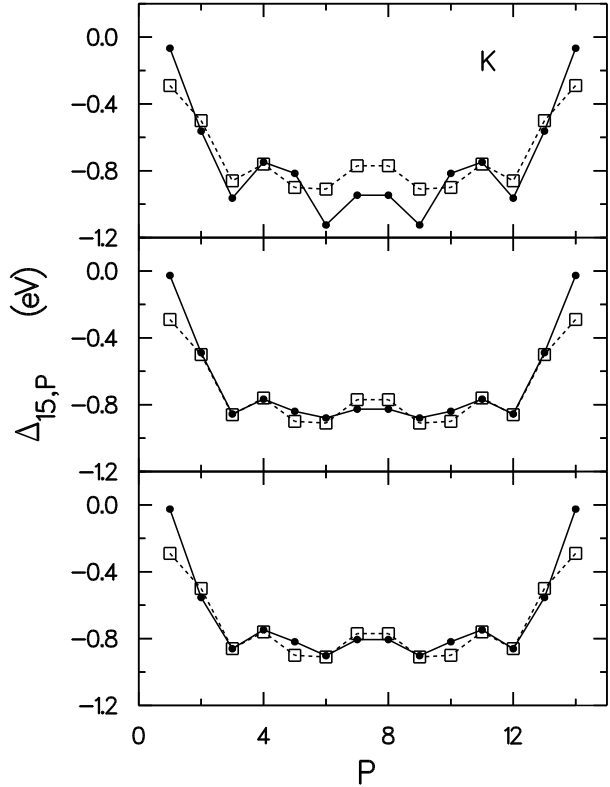}
\caption{
\small{
Fission dissociation energies, $\Delta_{15,P}$ [see Eq. (\protect\ref{delt})], 
for the doubly cationic K$^{2+}_{15}$ cluster 
as a function of the fission channels $P$.
Solid dots: Theoretical results derived from the SE-SCM method.
Open squares: Experimental measurements from Ref.\protect\ 
\protect\onlinecite{brec3}.
Top panel: The spherical model compared to experimental data.
Middle panel: The spheroidal model compared to experimental data.
Lower panel: The ellipsoidal model compared to experimental data.
}}
\end{figure}
for the cases of parent clusters having $N=26$, 23, 18, and 15 potassium atoms,
and compared them with experimental results \cite{brec3}.
The theoretical calculations compared to the 
experimental results are displayed in Figs.\ $(5 - 8)$ for 
$N=26$, 23, 18, 15, respectively. Again, while consideration of spheroidal 
shapes improves greatly the agreement between theory and experiment over the
spherical model, fully detailed correspondence is achieved only upon allowing 
for triaxial-shape deformations  (notice the improvement
in the range $P=12-14$ for $N=26$, and in the range $P=10-13$ for $N=23$).
In the cases $N=18$ and $N=15$ (Fig.\ 7 and Fig.\ 8), the biaxial and
triaxial results are essentially identical, since no 
fragment with more than nine electrons is involved. We note that the magic 
fragments K$^+_3$ and K$^+_9$ correspond always to strong minima,
and that for $N=18$ the channel associated with the double magic fragments 
(K$^+_9$, K$^+_9$) is clearly the favored one over the other magic channel 
with K$^+_3$, in agreement with the experimental analysis.

\begin{figure}[t]
\centering\epsfig{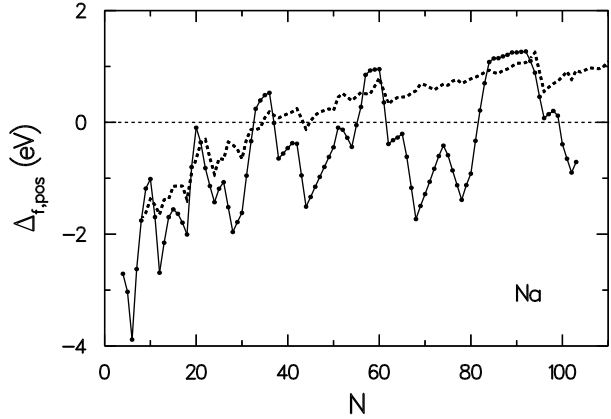}
\caption{
\small{
Solid dots: LDA-SCM results for the 
dissociation energies $\Delta^{\text{pos}}_f$ for the most favorable 
fission channel for doubly charged cationic parents Na$_N^{2+}$ when the 
spherical jellium is used. The influence of triaxial deformation effects 
(calculated with the SE-SCM approach) is shown by the thick dashed line.
}}
\end{figure}
\begin{figure}[ht]
\centering\epsfig{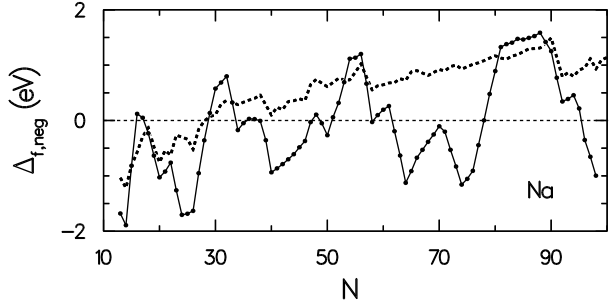}
\caption{
\small{
Solid dots: LDA-SCM results for the 
dissociation energies $\Delta^{\text{neg}}_f$ for the most favorable 
fission channel for doubly charged anionic parents Na$_N^{2-}$ when the 
spherical jellium is used. The influence of triaxial deformation effects 
(calculated with the SE-SCM approach) is shown by the thick dashed line.
}}
\end{figure}
Finally, we carried out calculations of dissociation energies,
$\Delta_f^{\text{pos}}$ and $\Delta_f^{\text{neg}}$, of the most favored
fission channels over the whole range up to $N=100$ atoms for the cases of
doubly charged cationic and anionic sodium clusters, respectively. 
The triaxial results compared to the spherical-jellia calculations according to
the LDA-SCM method \cite{yl1} are displayed in Fig.\ 9 and 
Fig.\ 10. In both cases, the main difference from the spherical jellium
is a strong suppression of the local minima, indicating that the critical
size for exothermic fission is significantly smaller than $N=100$
(about $N=30$), as indeed 
has been observed experimentally for hot cationic alkali-metal clusters
\cite{brec3} (the spherical-jellium results clearly are not compatible with the
emergence of such experimental critical sizes in the size range $N \leq 100$).

\subsection{Electronic shell effects in fission barriers and fission
dynamics of metal clusters}

In this section, we focus our discussion on recent trends in studies of 
binary fission processes in doubly charged metal clusters.

\subsubsection{Molecular-dynamics studies of fission}

Before discussing applications of the SE-SCM (and variants thereof) 
to the description of cluster fission, we note that for atomic and
molecular clusters microscopic descriptions of energetics and dynamics
of fission processes, based on modern electronic structure calculations
in conjunction with molecular dynamics simulations (where the 
classical trajectories of the ions, moving on the concurrently
calculated Born-Oppenheimer (BO) 
electronic potential energy surface, are obtained via 
integration of the Newtonian equations of motion), are possible and have
been performed \cite{barn,barn2} using the 
BO-local-spin-density(-LSD)-functional-MD method \cite{bl}. 
Such calculations, using norm-conserving non-local
pseudopotentials and self-consistent solutions of the KS-LSD equations 
\cite{barn,barn2}, applied to small sodium \cite{barn} and potassium 
\cite{barn2} clusters, revealed several important trends (Figs.\ $11-13$):
\begin{figure}[t]
\centering\epsfig{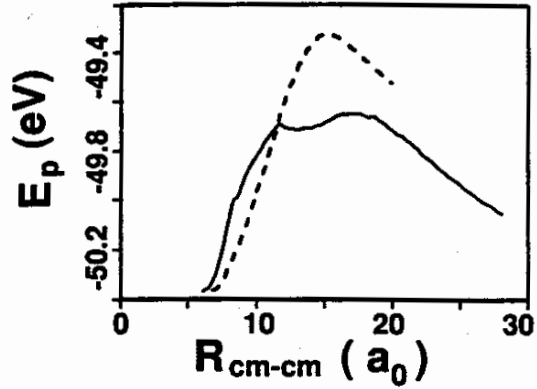}
\caption{
\small{
Molecular dynamics results for the
potential energy vs distance (in atomic units) between the centers of mass
for the fragmentation of Na$_{10}^{2+}$ into Na$_7^+$ and Na$_3^+$ (solid)
and Na$_9^+$ and Na$^+$ (dashed), obtained via constrained minimization of the
LSD ground-state energy of the system \protect\cite{barn}.
}}
\end{figure}
\begin{figure}[t]
\centering\epsfig{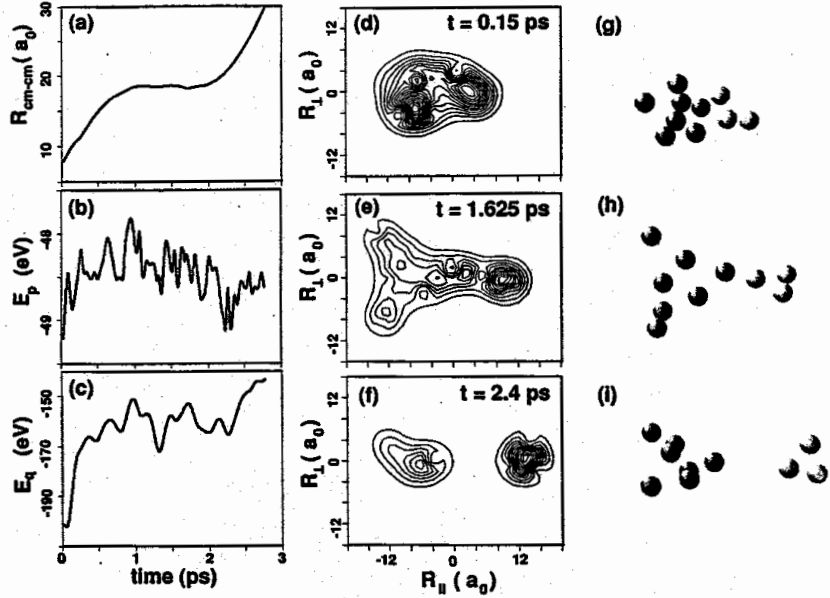}
\caption{
\small{
Fragmentation dynamics of Na$_{10}^{2+}$ from first-principles 
Born-Oppenheimer Local-spin-density functional Molecular-dynamics
simulations \protect\cite{barn}.
(a)-(c) Center-of-mass distance between the eventual fission
products ($R_{\text{c.m.-c.m.}}$), total potential energy ($E_p$), and the 
electronic contribution $E_q$ to $E_p$, vs time. (d)-(f) Contours of the total
electronic charge distribution at selected times calculated in the plane
containing the two centers of mass. The $R_{||}$ axis is parallel to
{\bf R}$_{\text{c.m.-c.m.}}$. (g)-(i) Cluster configurations for the times
given in (d)-(f). Dark and light balls represent ions in the large and
small fragments, respectively. Energy, distance, and time in units of eV, bohr
($a_0$), and ps, respectively.
}}
\end{figure}
\begin{figure}[p]
\centering\epsfig{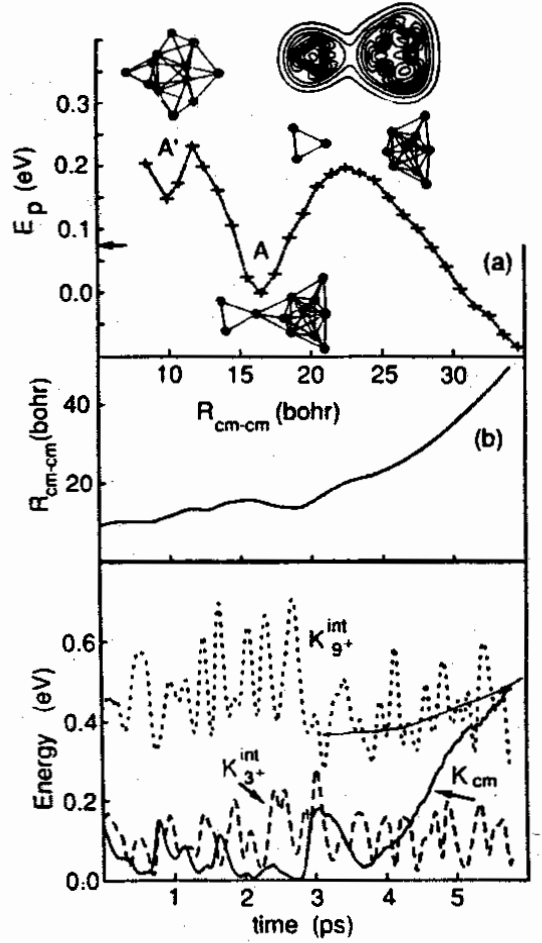}
\caption{
\small{
(a) Potential energy of K$_{12}^{2+}$ fissioning in the favorable channel
(K$_3^+ +$K$_9^+$) versus the inter-fragment distance 
$R_{\text{c.m.-c.m.}}$ obtained via
constrained minimization. The origin of the $E_p$ scale is set at the optimal
pre-barrier configuration (A). For large $R_{\text{c.m.-c.m.}}$, 
$E_p$=$-$0.9 eV,
i.e., $\Delta_3$. Included also are cluster configurations of K$_{12}^{2+}$
corresponding to: a compact isomer (A$^\prime$) 
(the energy of the optimal compact
isomer found is denoted by an arrow); the optimal bound configuration (A);
the structure on top of the exit-channel barrier for which contours of the 
total electronic charge density, $\rho$, are shown\protect\cite{barn2}. 
\protect\\
(b) Time evolution of $R_{\text{c.m.-c.m.}}$, the internal vibrational kinetic
energies of the fragments (K$_{3^+}^{\text{int}}$ and 
K$_{9^+}^{\text{int}}$) and the
sum of the fragments translational kinetic energies ($K_{\text{cm}}$) 
obtained via a
BO-LSD-MD simulation starting from ionization ($t=0$) of a K$_{12}^+$ cluster
at 500 K. A line is drawn in K$_{9^+}^{\text{int}}$ (for $t \geq 3$ ps) to
guide the eye, illustrating heating of the internal vibrational degrees
of freedom of the departing fragment.
}}
\end{figure}
(i) The energetically favorable fission channel for such doubly-charged
clusters is the asymmetric one, M$_N^{2+} \rightarrow$ M$_{N-3}^{+} +
$ M$_3^+$, containing a "magic" daughter M$_3^+$ (M $=$ Na, K), i.e.,
$\Delta_{N,P} = E($M$_{N-P}^+) + E($M$_P^+) - E($M$_N^{2+})$ is smallest
for $P=3$; (ii) Fission of clusters with $N \geq N_b^{2+}$, where
$N_b^{2+}=7$, involves barriers, whose magnitudes reflect the closed-shell
stability of the parent cluster (i.e., $E_b$ for
$N=10$ is particularly high), exhibiting a double-humped
barrier shape [see, Figs.\ 11 and 13(a)]; (iii) The eventual fission products
may be distinguishable (i.e., preformed) already 
at a rather early stage of the fission process
(on the top of the exit barrier for Na$_{10}^{2+}$, see Fig.\ 12, or prior
to the exit barrier for K$_{12}^{2+}$, see Fig.\ 13), and the electronic
binding between the two fragments is long-range in nature; (iv) The kinetic
energy release ${\cal E}_r$ in the favorable channel obtained via dynamic
simulations was found to be given by ${\cal E}_r \approx E_b + |\Delta_{N,3}|$,
and the results are in correspondence with experimental measurements
\cite{barn2} for K$_N^{2+}$ ($5 \leq N \leq 12$). Furthermore, in
agreement with experimental findings, the emerging fragments are 
vibrationally excited, with the heating of the internal nuclear degrees of
freedom of the fission products in the exit channel originating from dynamical
conversion of potential into internal kinetic energy 
[see, K$_{9^+}^{\text{int}}$ in Fig.\ 13(b)].

\subsubsection{SE-SCM interpretation of fissioning processes}

The method we adopt in this section for further studying metal-cluster fission
is the SE-SCM described in section II.B.6 (see also Ref.\ \cite{yl7}).

As discussed above (see section II.B),
in the SE-SCM method we need to introduce appropriate empirical potentials.
As will become apparent from our results, one-center potentials (like the 
one-center modified, anisotropic harmonic oscillator)
are not adequate for describing shell effects in the
fission of small metal clusters; rather, a two-center potential is required.
Indeed, the empirical potentials should be able to simulate the fragmentation 
of the initial parent cluster towards a variety
of asymptotic daughter-cluster shapes, e.g., two spheres in the
case of double magic fragments, a sphere and a spheroid in the case of a 
single magic fragment, or two spheroids in a more general case. 
In the case of metal clusters, asymmetric channels are most common,
and thus a meaningful and flexible description of the asymmetry is of primary 
concern. We found \cite{yl7}
that such a required degree of flexibility can be provided via
the shape parametrization of the asymmetric two-center-oscillator shell model 
(ATCOSM) introduced earlier in nuclear fission \cite{grei} (see section 
II.B.6).

In addition to the present shape parametrization \cite{yl7}, other
two-center shape parametrizations [mainly in connection with 
KS-LDA jellium calculations] have been used \cite{garc1,koiz1,rigo} in
studies of metal cluster fission.
They can be grouped into two categories, namely, the two-intersected-spheres 
jellium \cite{garc1,note222}, and the variable-necking-in parametrizations
\cite{koiz1,rigo}. In the latter group, Ref.\ \cite{koiz1}
accounts for various necking-in situations by using the "funny-hills" 
parametrization \cite{note333}, while Ref.\ \cite{rigo} describes the 
necking-in by connecting two spheres smoothly through a quadratic surface. The 
limitation of these other parametrizations is that they are not flexible 
enough 
to account for the majority of the effects generated by the shell structure of
the parent and daughters, which in general do not have spherical, but deformed 
(independently from each other), shapes. An example is offered by the case of 
the parent Na$_{18}^{2+}$, which has a metastable oblate ground state, and 
thus cannot be described by any one of the above parametrizations. We wish to
emphasize again that one of the conclusions of the present work is that 
the shell structures of the (independently deformed) parent and daughters are 
the dominant factors specifying the fission barriers, and thus 
parametrizations \cite{garc1,koiz1,rigo} 
with restricted final fragment (or parent) shapes 
are deficient in accounting for some of the most important features governing 
metal-cluster fission.

As a demonstration of our method,
we present results for two different parents, namely Na$_{10}^{2+}$ and
Na$_{18}^{2+}$.

\begin{figure}[p]
\centering\epsfig{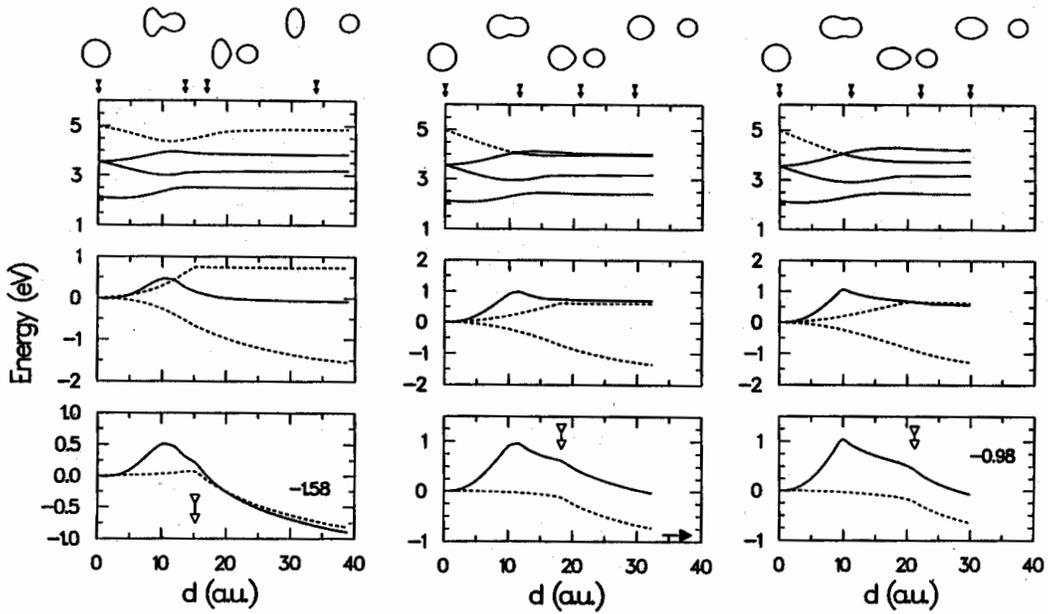}
\caption{
\small{
ATCOSM results for the asymmetric channel Na$_{10}^{2+}$ $\rightarrow$
Na$_7^+ +$Na$_3^+$. The final configuration of Na$_3^+$ is spherical.
For the heavier fragment Na$_7^+$, we present results associated
with three different final shape configurations, namely, oblate 
[(o,s); left], spherical [(s,s); middle], and prolate [(p,s); right]. 
The ratio of shorter over longer
axis is 0.555 for the oblate case and 0.75 for the prolate case. \protect\\
Bottom panel: LDM energy (surface plus Coulomb, dashed curve) 
and total potential energy (LDM plus
shell corrections, solid curve) as a function of fragment separation $d$.
The empty vertical arrow marks the scission point. 
The zero of energy is taken at $d=0$.
A number ($-$1.58 eV or $-$0.98 eV), or a horizontal solid arrow, 
denotes the corresponding dissociation energy.
\protect\\
Middle panel: Shell-correction contribution (solid curve),
surface contribution (upper dashed curve), and Coulomb contribution
(lower dashed curve) to the total energy, as a function of fragment
separation $d$. \protect \\
Top panel: Single-particle spectra as a function of fragment separation
$d$. The occupied (fully or partially) levels are denoted with solid lines. 
The unoccupied levels are denoted with dashed lines. \protect\\
On top of the figure, four snapshots of the evolving cluster shapes are
displayed. The solid vertical arrows mark the corresponding fragment
separations.
Observe that the doorway molecular configurations correspond to the second
snapshot from the left.
Notice the change in energy scale for the middle and bottom panels, as one
passes from (o,s) to (s,s) and (p,s) final configurations.
}}
\end{figure}
Fig.\ 14 presents results for the channel 
Na$_{10}^{2+}$ $\rightarrow$ Na$_{7}^+$ $+$ Na$_3^+$ for three different cases,
namely, when the larger fragment Na$_{7}^+$ is oblate (left column),
spherical (middle column), and prolate (right column). From our one-center
analysis, we find as expected that Na$_{7}^+$ (with six electrons) has an 
oblate global minimum and a higher in energy prolate local minimum. In the 
two-center analysis,
we have calculated the fission pathways so that the emerging fragments 
correspond to possible deformed one-center minima. 
It is apparent that the most favored channel (i.e., having the lowest barrier,
see the solid line in the bottom panels) 
will yield an oblate Na$_{7}^+$ (left column in Fig.\ 14), 
in agreement with the expectations from the 
one-center energetics analysis. 

The middle panels exhibit the decomposition of the total barrier into the three
components of surface, Coulomb, and shell-correction terms [see Eq. 
(\ref{et})], which are denoted by an upper dashed curve, a lower dashed curve,
and a solid line, respectively. The total LDM contribution (surface plus
Coulomb) is also exhibited at the bottom panels (dashed lines).

It can be seen that the LDM barrier is either absent or very small, and 
that the total barrier is due almost exclusively to electronic shell effects. 
The total barrier has a double-humped structure, with the outer hump
corresponding to the LDM saddle point, which also happens to be the scission
point (indicated by an empty vertical arrow). The inner hump coincides with
the peak of the shell-effect term, and is associated with the
rearrangement of single-particle levels from the initial spherical
parent to a molecular configuration resembling a Na$_7^+$ attached to
a Na$_3^+$. 
Such molecular configurations (discovered earlier in first-principles
MD simulations \cite{barn,barn2} 
of fission of charged metal clusters, as well as in studies of fusion of 
neutral clusters \cite{knos}) are a natural precursor towards full fragment
separation and complete fission, and naturally they give rise to the 
notion of preformation of the emerging fragments \cite{barn,barn2}.

\begin{figure}[t]
\centering\epsfig{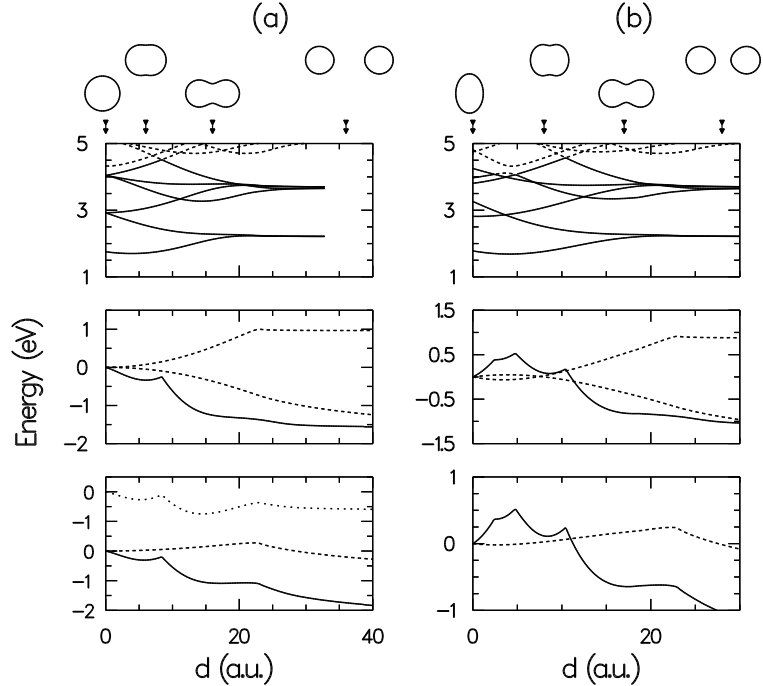}
\caption{
\small{
ATCOSM results for the symmetric channel Na$_{18}^{2+}$ $\rightarrow$
2Na$_9^+$, when the initial parent shape is assumed (a) spherical, and
(b) oblate (with a shorter over longer axis ratio equal to 0.699).
Panel distribution and other notations and conventions are the same as in 
Fig.\ 14. The top dotted line in the bottom panel of (a) represents the total
energy without the Coulomb contribution.
Observe that the doorway molecular configurations correspond to the third
snapshot from the left.
Notice that the zero of all energies is taken at $d=0$.
}}
\end{figure}
Fig.\ 15(a) displays the ATCOSM results for the symmetric channel 
Na$_{18}^{2+}$ $\rightarrow$ 2Na$_9^+$, when, for illustrative purposes, the 
parent is assumed to be spherical at $d=0$ 
(this channel is favored compared to that of the trimer \cite{yl7}, 
both from energetics and barrier considerations;
for small clusters, this is the only case where a channel other than that of
the trimer is the most favored one).
The top panel of Fig.\ 15(a) describes the evolution of the
single-particle spectra. The spherical ordering $1s$, $1p$, $1d$, $2s$, etc.,
for the parent at $d=0$ is clearly discernible. With increasing separation
distance, the levels exhibit several crossings, and, after the scission point,
they naturally regroup to a new ordering associated with the
spherical Na$_9^+$ products (at the end of the fission process, the levels are
doubly degenerate compared
to the initial configuration, since there are two Na$_9^+$ fragments). It is
seen that the ATCOSM leads to an oscillator energy 
(i.e., the gap between two populated major shells exhibited at the right end 
of the figure) of 1.47 eV for each Na$_9^+$ fragment in agreement with the 
value expected from the one-center model [the $1s$ state of Na$_9^+$ lies at 
2.21 eV; in the case of the initial spherical Na$_{18}^{2+}$ ($d=0$), 
the oscillator energy corresponding to the gap between major shells is 1.17 eV,
and the corresponding $1s$ state lies at 1.75 eV].

From the middle panel of Fig.\ 15(a), we observe that the shell-correction 
(solid line) contributes a net gain in energy of about 1.6 eV upon dissociation
into two Na$_9^+$ fragments.  This gain is larger than the increase in energy 
(i.e., positive energy change) due to the surface term, which saturates at a 
value of about 1 eV after the scission point at $d \approx 23$ a.u.
The total energy is displayed in the bottom panel of
Fig.\ 15(a) (solid line) along with the LDM barrier (dashed line). Even though
distorted (when compared to the cases of Fig.\ 14), 
the total barrier still exhibits a two-peak structure, the inner
peak arising from the hump in the shell correction, and the outer peak
arising from the point of saturation of the surface term (this last
point coincides again with the scission point, as well as with the saddle of the
LDM barrier). An inner local minimum is located at $d \approx $ 8 a.u., and
corresponds to a compact prolate shape of the parent [see second drawing 
from the left at the top
of Fig.\ 15(a)], while a second deeper minimum appears at $d \approx$ 18 a.u.,
corresponding to a superdeformed shape of a molecular configuration of two 
Na$_9^+$ clusters tied up together [preformation of fragments, 
see third drawing from the left at the top of Fig.\ 15(a)]. 
The inner barrier separating the compact
prolate configuration from the superdeformed molecular configuration arises
from the rearrangement of the single-particle levels during the transition from
the initially assumed spherical Na$_{18}^{2+}$ configuration to that of the 
supermolecule
Na$_9^+$+Na$_9^+$. We note that the barrier separating the molecular 
configuration from complete fission is very weak being less than 0.1 eV.

The top dotted line at the bottom panel displays
the total energy in the case when the Coulomb contribution is neglected. This
curve mimics the total energy for the fusion of two neutral Na$_8$ clusters,
namely the total energy for the reaction 2Na$_{8}$ $\rightarrow$ Na$_{16}$.
Overall, we find good agreement with a KS-LDA
calculation for this fusion process (see Fig.\ 1 of Ref.\
\cite{knos}).
We further note that the superdeformed minimum for the neutral Na$_{16}$
cluster is deeper than that in the case of the doubly charged Na$_{18}^{2+}$
cluster. Naturally,
this is due to the absence of the Coulomb term. 
                                  
The natural way for producing experimentally the metastable Na$_{18}^{2+}$ 
cluster is by ionization of the stable singly-charged Na$_{18}^+$ cluster. 
Since this latter cluster contains
seventeen electrons and has a deformed oblate ground state [40a],
it is 
not likely that the initial configuration of Na$_{18}^{2+}$ will be spherical
or prolate as was assumed for illustration purposes in Fig.\ 15(a). 
Most likely, the initial configuration
for Na$_{18}^{2+}$ will be that of the oblate Na$_{18}^+$. To study the effect
that such an oblate initial configuration has on the fission barrier, 
we display
in Fig.\ 15(b) ATCOSM results for the pathway for the symmetric fission 
channel, starting from an oblate shape of Na$_{18}^{2+}$,
proceeding to a compact prolate shape, and then to full separation
between the fragments 
via a superdeformed molecular configuration. We observe that 
additional potential humps (in the range 2 a.u. $\leq d \leq$ 6 a.u.),
associated with the shape transition from the oblate to the compact prolate
shape, do develop. Concerning the total energies, the additional
innermost humps result in the emergence of a significant fission barrier of 
about 0.52 eV for the favored symmetric channel [see $ d \approx 5$ a.u. in 
Fig.\ 15(b)]. 

From the above analysis, we
conclude that considerations of the energy pathways leading from the parent to
preformation configurations (i.e., the inner-barrier hump, or humps) together
with the subsequent separation processes are most important for proper
elucidation of the mechanisms of metal-cluster fission processes. 
This corroborates earlier results obtained via first-principles
MD simulations \cite{barn,barn2} pertaining to the 
energetics and dynamical 
evolution of fission processes, and emphasizes that focusing exclusively 
\cite{garc1,rigo} on the separation process between the preformed state and 
the ultimate fission products provides a rather incomplete description of 
fission phenomena in metal clusters. It is anticipated that, with the use of
emerging fast spectroscopies \cite{wost}, experimental probing of the 
detailed dynamics of such fission processes could be achieved.

\section{Influence of electronic entropy on shell effects}

In the previous sections, we showed that consideration of triaxial
(ellipsoidal) shapes in the framework of the SCM leads to overall substantial
systematic improvement in the agreement between theory and experimental 
observations pertaining to the major and the fine structure of the 
size-evolutionary patterns associated with the energetics of fragmentation 
processes (monomer/dimer dissociation energies and fission energetics) and 
ionization. 

The theoretical methods and discussion of deformation effects 
in the previous sections were restricted to zero temperature. However, the 
experiments are necessarily made with clusters at finite temperatures, 
a fact that strongly motivates the development of finite-temperature 
theoretical approaches.

Due to the difficulty of the subject, to date only a few finite-temperature 
theoretical studies of metal clusters have been performed. In this section, 
we discuss briefly some of the conclusions of a recent SCM study \cite{yl5}
regarding the importance of thermal effects. The theoretical details pertaining
to this finite-temperature (FT) $-$SE$-$SCM will not be elaborated here, 
but they can be found in the aforementioned reference.

\begin{figure}[t]
\centering\epsfig{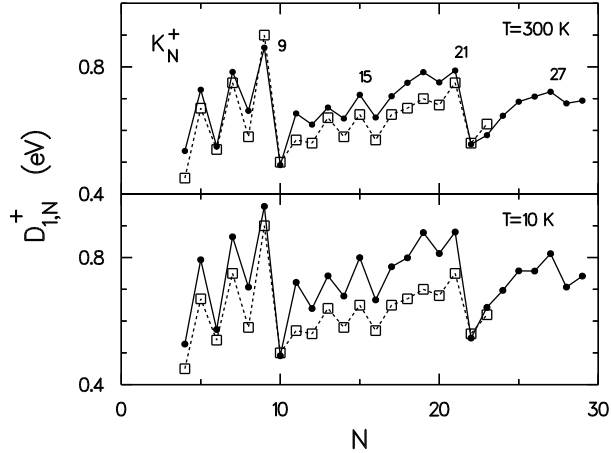}
\caption{
\small{
Monomer separation energies of K$_N^+$ clusters at two temperatures, 
$T=10$ K, and 300 K.
Solid dots: Theoretical FT-SE-SCM results. Open squares: experimental 
measurements \protect\cite{brec1}. 
To facilitate comparison, the SE-SCM results 
at the higher temperature have been shifted by 0.07 eV, so that the 
theoretical curves at both temperatures refer to the same point at $N=10$.
}}
\end{figure}
The main conclusion of Ref.\ \cite{yl5} was that, in conjunction with
deformation effects, electronic-entropy
effects in the size-evolutionary patterns of relatively small 
(as small as 20 atoms) simple-metal clusters become prominent already at
moderate temperatures. At smaller sizes, electronic-entropy effects are less
prominent, but they can still be discernible.
As an example, we present in Fig.\ 16
the monomer separation
energies of K$_N^+$ clusters for two temperatures ($T=10$ K and $T=300$ K),
along with the available experimental measurements \cite{brec1} (open
squares) in the size range $N=$ 4 $-$ 23.
First notice that the $T=10$ K results are practically indistinguishable
from the $T=0$ K results presented in Fig.\ 2.
Compared to the $T=10$ K results, the theoretical results at 
$T=300$ K are in better agreement with the experimental ones
due to an attenuation of the amplitude of the alternations
(e.g., notice  the favorable reduction in the size of the drops at
$N=9$, 15, and 21). This amplitude attenuation, however, is moderate, and 
it is remarkable that the $T=300$ K SCM results in this size range preserve in 
detail the same relative pattern as the $T=0$ K ones (in particular, 
the well-defined odd-even oscillations in the range $N=4-15$ and the
ascending quartet at $N=16-19$ followed by a dip at $N=20$).

\begin{figure}[t]
\centering\epsfig{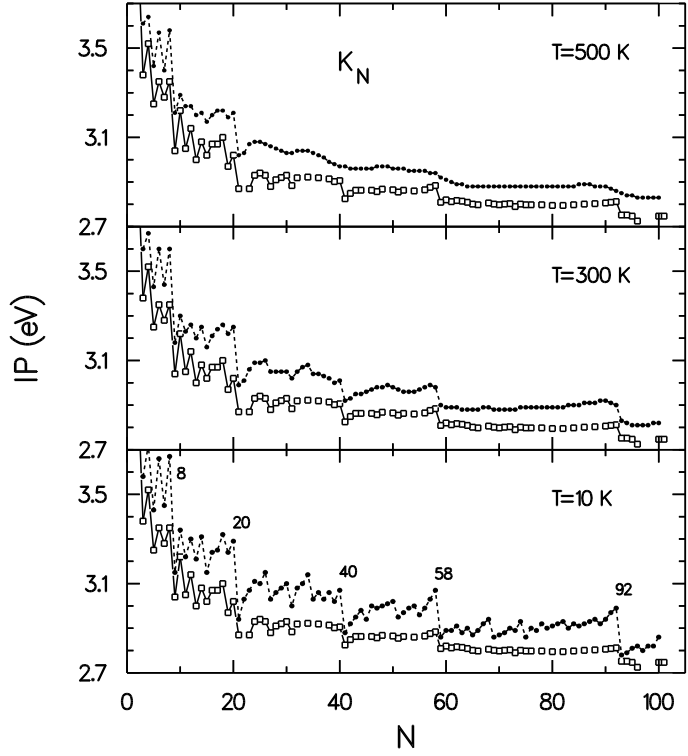}
\caption{
\small{
IPs of K$_N$ clusters at three temperatures, $T=10$ K, 300 K, and 500 K.
Solid dots: Theoretical FT-SE-SCM results. Open squares: experimental 
measurements \protect\cite{saun}.
}}
\end{figure}
As a further example, the theoretical IPs of K$_N$ clusters in the size range
$ 3 \leq N \leq 102$ for three temperatures, $T=10$ K, 300 K, and 500 K,
are displayed in Fig.\ 17, and are compared with the experimental measurements
\cite{saun} (open squares; the experimental uncertainties are 
0.06 eV for $N \leq 30$ and 0.03 eV for $N > 30$). 
As was the case with our earlier $T=0$ K results
\cite{yl3}, the $T=10$ K theoretical results exhibit the following 
two characteristics: (i) Above $N=21$, a 
pronounced fine structure between major-shell closures which is not present in
the experimental measurements; (ii) Steps at the major-shell closures which
are much larger than the experimental ones, i.e., three-to-five times 
for $N=40$, 58, and 92, and two-to-three times for $N=8$ and 20
[this needs to be contrasted to the experimental IPs for cold Na$_N$ clusters,
which are in overall good agreement with out $T=0$ K SE-SCM
results regarding both characteristics (see Fig.\ 1)].

The agreement between theory and experiment is significantly improved at
$T=300$ K. Indeed, in comparison with the lower-temperature
calculations, the $T=300$ K
results exhibit the following remarkable changes: (i) Above $N=21$, the
previously sharp fine-structure features are smeared out, and as a 
result, the theoretical curve follows closely the mild modulations of the  
experimental profile. In the size range $N=21-34$, three rounded, 
hump-like formations (ending to the right at the subshell closures at $N=26$,
30, and 34) survive in very good agreement with the experiment (the sizes of 
the drops at $N=26$, 30 and 34 are comparable to the experimental ones
\cite{note30}); (ii) The sizes of the IP drops at $N=20$, 40, 58, and 92
are reduced drastically and are now comparable to the experimental ones.
In the size range $N \leq 20$, the modifications are not as dramatic. Indeed,
one can clearly see that the pattern of odd-even alternations
remains well defined, but with a moderate attenuation in amplitude, again 
in excellent agreement with the experimental observation. 

For $T=500$ K, the smearing out of the shell structure associated with the
calculated results  
progresses even further, obliterating the agreement between theory and 
experiment. Specifically, the steps at the subshell closures at $N=26$ and
30, as well as at the major-shell closures at $N=40$, 58, and 92 are rounded
and smeared out over several clusters
(an analogous behavior has been observed in the logarithmic abundance spectra
of hot, singly cationic, copper, silver, and gold clusters \cite{kata}).
At the same time, however, the odd-even alternation remains well defined
for $N \leq 8$. We further notice that, while some residue of fine structure 
survives in the range $N=9-15$, the odd-even alternations there are 
essentially absent (certain experimental measurements \cite{kapp} of the IPs 
of hot Na$_N$ clusters appear to conform to this trend).

The influence of the electronic entropy on the height of fission barriers
has not
been studied as yet, but it will undoubtedly be the subject of future
research in metal-cluster physics. In any case, 
based on the results of this section, it is natural to conjecture 
that electronic-entropy effects will tend to quench the barrier heights, 
especially in the case of larger multiply charged clusters.

\section{Summary}

In this chapter, we have elucidated certain issues pertaining to evaporation 
and fission processes of metallic clusters, focusing on electronic
shell effects and their importance in determining the energetics, structure,
pathways, and dynamical mechanisms of dissociation and fragmentation in these 
systems, and have outlined and demonstrated various theoretical approaches 
currently used in investigations of cluster fragmentation phenomena, ranging 
from microscopic first-principles
electronic structure calculations coupled with molecular dynamics simulations
to adaptation of more phenomenological in nature models originated in 
studies of atomic nuclei. In this respect, a recurrent theme in this 
exposition has been the crucial importance of 
deformation and electronic-entropy (temperature) effects, as well as
their treatment with the help of shell correction methods. 

By drawing analogies, as well as differences,
between certain aspects of nuclear fission and nuclear radioactivity 
phenomena and atomic
(metallic) cluster fission processes, we have attempted to provide a unifying 
conceptual framework for discussion of the physical principles
underlying modes of cluster fission (i.e., importance of deformations, 
shell effects originating from
fragments and parent, asymmetric and symmetric fission, single and 
double-humped barriers, fissioning cluster shapes, and dynamical aspects, such
as the time-scale of fission processes, kinetic energy release, and dynamical
energy redistribution among the fission products).

We conclude by commenting on some experimental and theoretical issues in 
cluster fission which remain as future challenges (limiting ourselves to
metallic clusters). These include: fission dynamics of multiply charged large
metal clusters \cite{mart,brec11,mart11};
systematic investigations of temperature effects on modes
of cluster fission, and ternary, and higher multi-fragmentation processes;
time-resolved spectroscopy of fission processes and of fission isomers;
spin effects in fission; tunneling processes and corresponding life-times
in sub-barrier fission modes of clusters of light elements, e.g., lithium;
and fission processes of non-simple metal clusters.\\
~~~~~~~~~~~~~~\\
~~~~~~~~~~~~~~\\
~~~~~~~~~~~~~~\\
This research was supported by a grant from the U.S. Department of Energy
(Grant No. FG05-86ER45234). Calculations were performed on CRAY computers
at the Supercomputer Center at Livermore, California, and the Georgia
Institute of Technology Center for Computational Materials Science.


\begin{references}
\bibitem{note11}
For monovalent elements (Na, K, Cu, etc.),
$N$ is also equal to the number $N_e$ of delocalized valence electrons.
\bibitem{note1}
Here, we consider clusters of monovalent elements (Na, K, Cu, etc.).
For polyvalent elements, $N$ in Eq.\ (\ref{npar}) must be replaced by $Nv$,
where $v$ is the valency.
\bibitem{note43}
This meaning of the variable $\rho$ should not be confused with the meaning
of $\rho$ as a particle density in a density functional (see section II.A).
\bibitem{note222}
The two-intersected-spheres jellium has also been used for describing 
the fusion of two neutral magic clusters (see Ref.\ \cite{knos}).
\bibitem{note333}
In this three-variables parametrization, the $B$ parameter controls the
necking-in, the $C$ parameter controls the distance, and the $\alpha$
parameter controls the asymmetry, leaving no freedom for the shapes
of the parent or the emerging fragments to be varied. 
In particular, both fragments remain 
simultaneously either prolatelike or oblatelike, while final spherical
shapes are excluded altogether. The weaknesses of the "funny hills"
parametrization with respect to metal-cluster fission have been discussed
in Ref.\ \cite{mart11}.
\bibitem{note30}
Notice that experimental measurements at $N=33$ and 35 have not been obtained
(see Ref.\ \cite{saun}).
\end{references}
\end{document}